\documentclass{ws-mpla}

\usepackage[super]{cite}
\usepackage{xcolor}
\usepackage[verbose,hypertexnames=false]{hyperref}
\hypersetup{colorlinks=false,allbordercolors=blue,pdfborderstyle={/S/U/W 1}}
\begin{document}

\markboth{Saleev V. A., Shilyaev K. K.}{Prompt production of $J/\psi$ in the soft gluon resummation approach using the ICEM}

%%%%%%%%%%%%%%%%%%%%% Publisher's Area please ignore %%%%%%%%%%%%%%
\catchline{}{}{}{}{}
%%%%%%%%%%%%%%%%%%%%%%%%%%%%%%%%%%%%%%%%%%%%%%%%%%%%%%%%%%%%%%%%%%%

\title{Prompt production of $J/\psi$  in the soft gluon resummation approach using the ICEM}

\author{Vladimir Saleev}

\address{Samara National Research University, Moskovskoe Shosse, 34, 443086, Samara, Russia\\
Joint Institute for Nuclear Research, Dubna, 141980, Russia\\
saleev.vladimir@gmail.com}

\author{Kirill Shilyaev}

\address{Samara National Research University, Moskovskoe Shosse, 34, 443086, Samara, Russia\\
kirill.k.shilyaev@gmail.com}

\maketitle

\begin{history}
\received{(Day Month Year)}
\revised{(Day Month Year)}
\accepted{(Day Month Year)}
\published{(Day Month Year)}
\end{history}

\begin{abstract}
In the article, we study prompt $J/\psi$ production at small transverse momentum within the Transverse Momentum Dependent (TMD) factorization. The Soft Gluon Resummation (SGR) approach is used for modelling of TMD PDFs. The hadronization process of heavy charm quarks is described with the Improved Color Evaporation model (ICEM). In order to obtain spectra of $J/\psi$ at the arbitrary transverse momenta, fixed-order calculations within the collinear parton model (CPM) are also applied and further matched with the TMD calculations using the Inverse-Error Weighting (InEW) scheme. We present results of calculations for a wide range of collision energy with the fitting of the relevant nonperturbative parameter of the ICEM. Predictions for the SPD NICA experiment are also provided alongside comparison with the result of calculation using nonrelativistic QCD (NRQCD) framework.
\end{abstract}


\ccode{PACS Nos.: 13.60.Le, 14.40.Gx}

\section{Introduction}	

Transverse Momentum Dependent (TMD) factorization~\cite{Collins:2011zzd, Boussarie:2023izj} is an approach which provides access to the parton distribution functions (PDFs) at the kinematic domain where the final state of hadron collisions is produced with small transverse momentum with respect to hard scale of the process, $p_T^{} \ll Q$. Various final states are appropriate to study TMD PDFs: the Drell-Yan pair~\cite{Collins:1984kg, Moos:2023yfa, Bacchetta:2019sam, BermudezMartinez:2020tys, Ma:2013aca} and Higgs boson~\cite{Kauffman:1991jt, Boer:2014tka, Sun:2016kkh} production, processes that include weak interactions such as $W/Z$ bosons production,~\cite{Bacchetta:2018lna, Sun:2013hua, Dooling:2014kia} light meson production in SIDIS processes,~\cite{Boglione:2014oea, Bacchetta:2019qkv, Scimemi:2019cmh} production of heavy quarkonium~\cite{Ma:2012hh, Echevarria:2019ynx} and pairs of quarkonia.~\cite{Zhang:2014vmh, Lansberg:2017dzg, Scarpa:2019fol} All mentioned processes represent production of colorless or color-singlet (like in case of heavy quarkonium pairs) states, therefore the TMD factorization theorem, which allows to separate nonperturbative and hard scattering parts, can be proven rigorously for them. In contrast, single $J/\psi$-meson production can pass via color-octet states, so the radiation and exchange of final soft gluons occur and the TMD factorization breaks in such processes.~\cite{Boussarie:2023izj} However, in the present study, we probe single $J/\psi$ production in unpolarized proton-proton collisions for TMD parton model (TMD PM) application in the domain $p_T^{} \ll M_{J/\psi}^{}$, where the mass of $J/\psi$ plays the role of the hard scale.

The TMD PDFs describe internal three-dimensional structure of protons in the momentum space as they include distributions over both longitudinal momentum fraction $x$ and transverse momentum $\vec{q}_T^{}$ of partons. Their evolution with factorization scale $\mu$ and rapidity variable $\zeta$ is described within the two-scale Collins-Soper equations.~\cite{Collins:1981va, Collins:1988ig} The gluon TMD PDFs have been studied worse than the quark distributions so the present and future studies involve their analysis, modelling and extraction in forthcoming experiments such as ones at SPD NICA~\cite{Arbuzov:2020cqg} and EIC.~\cite{Accardi:2012qut} In order to model and take into account scale-evolution of the TMD PDFs, we use the Soft Gluon Resummation (SGR) approach.~\cite{Collins:1984kg, Yuan:1991we, Sun:2012vc} It allows one to model the TMD PDFs reducing them to collinear distributions at some initial, low but still perturbative scale. Nevertheless, the phenomenological component of the modelling remains significant.

Generally speaking, the parton dynamics inside protons is described with parton correlators.~\cite{Boer:1997nt, Mulders:2000sh} In case of unpolarized incoming protons, correlators are parameterized with unpolarized parton distributions (unpolarized PDFs) and distributions that correspond to transversely polarized quarks (Boer-Mulders function, BM PDF) or linearly polarized gluons (a gluon analogue of BM PDF). The contribution of the BM PDF is relatively small as it is the next-to-leading order in the coupling constant $\alpha_s^{}$ than the unpolarized PDF,~\cite{Sun:2011iw} the BM PDFs contribution can be also considered as $(q_T^{2} / M_h^{2})$-correction to unpolarized PDFs contribution, $M_h^{}$ is the soft scale, which is about a proton mass. Moreover, taking into account of initial parton polarization slightly influence on the $p_T^{}$-spectrum curve and doesn't impact on the total cross section, therefore the Boer-Mulders effect is to be preferably found in spin-dependent observables such as polarization, angular modulation or spin asymmetries~\cite{Boer:1999mm, Boer:1997nt} whilst we focus on the unpolarized $J/\psi$ cross section in this article. Those are the reasons why we neglect BM PDF in our current analysis.

Another issue to be encountered in heavy quarkonium production is the process of hadronization of the heavy quark pair produced in a hard partonic subprocess. The well-known framework for hadronization is nonrelativistic QCD (NRQCD).~\cite{Bodwin:1994jh} Because the process of hadronization is of the nonperturbative nature, certain inclusion of phenomenological parameter sets can't be avoided. In the NRQCD, those are long-distance matrix elements (LDMEs), they are unique for each Fock state which makes analysis of heavy quarkonium production quite non-trivial, that we demonstrated in our recent article on charmonium production in the SGR framework.~\cite{Saleev:2025ryh} In this study, we use another hadronization approach, the Improved Color Evaporation model (ICEM).~\cite{Ma:2016exq} In the ICEM, cross section of heavy quarkonium production is described as cross section of heavy quark pair production averaged over a certain range of invariant mass of the final quark pair. All the nonperturbative details of the process are isolated in the single phenomenological parameter $F^{J/\psi}$ which was observed to depend on the collision energy $\sqrt{s}$ in the center-of-mass of incoming protons, though. This way, we obtain an instrument to test and verify our calculations in two hadronization frameworks.

The cross section in the domain of large transverse momentum $p_T^{} \gg M_{J/\psi}^{}$, which we also studied, is described with the standard fixed-order calculations in the collinear parton model (CPM) in the leading order (LO) approximation. For the domain of the moderate transverse momentum $p_T^{} \sim M_{J/\psi}^{}$, the Inverse-Error Weighting~\cite{Echevarria:2018qyi} (InEW) scheme was used for matching of the TMD PM and CPM results of calculations.

Applying the ICEM framework to the experimental data at various energies, we fitted $F^{J/\psi}$ to these data and approximated this dependence in order to have an opportunity to make predictions for the future SPD NICA experiment and to do a comparison of our predictions within the ICEM and NRQCD frameworks.

\section{Soft Gluon Resummation approach}

The momenta of the initial partons within protons are conventionally written in a form of the Sudakov decomposition of the light-cone four-momenta $p_{1,2}^{}$ of protons:
\begin{equation}
q_1^\mu = x_1^{} p_1^\mu + y_1^{} p_2^\mu + q_{1T}^\mu,~~~~~~~q_2^\mu = x_2^{} p_2^\mu + y_2^{} p_1^\mu + q_{2T}^\mu
\end{equation}
with momentum fractions $x_{i}^{}$ and $y_{i}^{}$, transverse momenta $q_{iT}^{}$ such that $q_{iT}^{2} = - \vec{q}_{iT}^{\,\,2}$. The gauge-invariance, i.e. the on-shell condition for the initial partons momenta $q_{i}^2 = 0$, demands the following expression $y_{i} = \vec{q}_{iT}^{\,\,2}/(sx_{i})$ for the fractions $y_{i}^{}$. In order to continue analysis in the leading approximation with resprect to the initial transverse momenta, in other words we neglect the kinematic power corrections of order $(q_{T}^{2}/Q^2)$ and higher, we assume that the fractions $y_{i}^{}$ approach to zero, therefore the momenta of initial partons are expressed as $q_{i}^{\mu} = x_{i}^{} p_{i}^{\mu} + q_{iT}^{\mu}$. Their components also acquire the simplified forms:
\begin{equation}
q_1^{\mu} \approx \left( \frac{x_1^{} \sqrt{s}}{2}, \vec{q}_{1T}^{}, \frac{x_1^{} \sqrt{s}}{2} \right),~~~q_2^{\mu} \approx \left( \frac{x_2^{} \sqrt{s}}{2}, \vec{q}_{2T}^{}, - \frac{x_2^{} \sqrt{s}}{2} \right),
\end{equation}
where $\vec{q}_{T}^{}$ are two-dimensional vectors of transverse momenta. 

The momenta of the partons within the parent protons are described with the parton correlators,~\cite{Boer:1997nt, Mulders:2000sh} the one for unpolarized protons is parameterized with the unpolarized parton distribution function $f_{1} (x, \vec{q}_{T}^{}, \mu_{}^{},\zeta)$ (unpolarized PDF) and the Boer-Mulders function for initial quarks (BM PDF) or its analogue for initial gluons, both are denoted as $h_{1}^{\perp} (x, \vec{q}_{T}^{}, \mu_{}^{}, \zeta)$. The contribution of the BM PDF to the transverse momentum spectrum is not expected to exceed $10-20\%$,~\cite{Bor:2022fga} that's why we neglect the BM PDF contribution in this work, as was also previously explained.

Theorem of the TMD factorization prescribes to write down expression for cross section as a convolution of hard-scale partonic subprocess cross section $d\hat{\sigma}$ and TMD parton distributions over momentum fractions $x_i^{}$ and initial transverse momenta $\vec{q}_{iT}^{}$ as follows:~\cite{Boussarie:2023izj}
\begin{equation}
d\sigma^{TMD} = \int dx_1^{} dx_2^{} \, d^2 q_{1T}^{} \,d^2 q_{2T}^{}\, f_1^{} (x_1^{}, \vec{q}_{1T}^{}, \mu_{}^{}, \zeta_1^{}) f_1^{} (x_2^{}, \vec{q}_{2T}^{}, \mu_{}^{}, \zeta_2^{})\, d\hat{\sigma},
\end{equation}
the partonic subprocess denotes the final charmed quark-antiquark pair production via gluon-gluon fusion $gg \rightarrow c\bar{c}$ or light quark-antiquark annihilation subprocess $q\bar{q} \rightarrow c\bar{c}$. We rewrite this expression in terms of the summed transverse momentum of $c\bar{c}$-pair $p_T^{c\bar{c}}$ and its invariant mass $M_{c\bar{c}}^{}$ to deal with the $2 \rightarrow 1$ partonic subprocess:
\begin{align}
d\sigma^{TMD} = \int dx_1^{}\, dx_2^{} & \, d^2 q_{1T}^{} \,d^2 q_{2T}^{}\, f_1^{} (x_1^{}, \vec{q}_{1T}^{}, \mu_{}^{}, \zeta_1^{}) f_1^{} (x_2^{}, \vec{q}_{2T}^{}, \mu_{}^{}, \zeta_2^{}) \times \nonumber\\ 
&\times\int dM_{c\bar{c}}^2\, d^4 p^{c\bar{c}} \, \hat{\sigma}(\hat{s}) \,\delta(\hat{s} - M_{c\bar{c}}^2)\, \delta^{(4)} (p_{c}^{} + p_{\bar{c}}^{} - p^{c\bar{c}}),
\end{align}
where $\hat{\sigma}(\hat{s})$ is a total cross section of the partonic subprocess and $\hat{s} = (q_1^{} + q_2^{})^2 = (p_{c}^{} + p_{\bar{c}}^{})^2$ is a Mandelstam variable. Further, the following expression for differential cross section can be derived after analytical evaluation of the integrations:
\begin{equation}
\frac{d\sigma^{TMD}}{dp_T^{c\bar{c}} \,dy_{c\bar{c}}^{} \, dM_{c\bar{c}}^2} = \frac{2\pi p_T^{c\bar{c}}}{s} \, \hat{\sigma}(\hat{s}) \, f_1^{} (x_1^{}, \vec{q}_{1T}^{}, \mu_{}^{}, \zeta_1^{}) \otimes_T^{} f_1^{} (x_2^{}, \vec{q}_{2T}^{}, \mu_{}^{}, \zeta_2^{})
\end{equation}
with the rapidity $y_{c\bar{c}}^{}$ of the $c\bar{c}$-pair and convolution of TMD PDFs over transverse momentum which reads
\begin{align}
f_1^{} (x_1^{}, &\vec{q}_{1T}^{}, \mu_{}^{}, \zeta_1^{}) \otimes_T^{}  f_1^{} (x_2^{}, \vec{q}_{2T}^{}, \mu_{}^{}, \zeta_2^{}) = \nonumber \\
& = \int d\vec{q}_{1T}^{} \, d\vec{q}_{2T}^{} f_1^{} (x_1^{}, \vec{q}_{1T}^{}, \mu_{}^{}, \zeta_1^{}) f_1^{} (x_2^{}, \vec{q}_{2T}^{}, \mu_{}^{}, \zeta_2^{}) \delta^{(2)} (\vec{q}_{1T}^{} + \vec{q}_{2T}^{} - \vec{p}_{T}^{\,c\bar{c}}).
\label{eq:conv}
\end{align}

In order to consider the evolution of TMD PDFs with the scales $\mu$ and $\zeta$, one should make the 2D Fourier transform of the distributions:~\cite{Collins:2011zzd}
\begin{equation}
\hat{f}_1^{}(x, \vec{b}_{T}^{}, \mu_{}^{}, \zeta) = \int d^2 q_{T}^{} \, e^{i \vec{q}_{T}^{} \vec{b}_{T}^{}} f_1^{}(x, \vec{q}_{T}^{}, \mu_{}^{}, \zeta),
\end{equation}
so that the transfer to the impact parameter $\vec{b}_{T}^{}$ space is performed. Within the Soft Gluon Resummation (SGR) approach, after the Fourier transform which turns the TMD PDFs convolution~(\ref{eq:conv}) into product, the solution for the Collins-Soper and Renormalization equations can be obtained in a multiplicative form of PDFs at the initial scale and evolution factor which is called perturbative Sudakov factor:~\cite{Collins:1981uk}
\begin{equation}
\hspace{-0.2cm} \hat{f}_1^{}(x_1^{}, b_{T}^{}, \mu, \zeta) \hat{f}_1^{}(x_2^{}, b_{T}^{}, \mu, \zeta)  = e^{-S_P^{} (\mu, \mu_{b}^{}, b_{T}^{})} \hat{f}_1^{}(x_1^{}, b_{T}^{}, \mu_{b}^{}, \mu_{b}^{2}) \hat{f}_1^{}(x_2^{}, b_{T}^{}, \mu_{b}^{}, \mu_{b}^{2}),
\end{equation}
where the following standard scale choice was made for simplicity: $\mu_0^{} = \sqrt{\zeta_0} = \mu_b^{}$ for initial scales and $\mu = \sqrt{\zeta}$ for the final scales. The Sudakov factor $S_P^{}$, which implements the perturbative evolution of TMD PDFs, is process-dependent, this is manifested in the coefficients $A(\mu')$ and $B(\mu')$ of the Sudakov factor expression:~\cite{Collins:1981uk, Aybat:2011zv}
\begin{equation}
S_P^{} (\mu, \mu_b^{}, b_T^{}) = \int\limits_{\mu_b^2}^{\mu^2} \frac{d\mu'^2}{\mu'^2} \left[ A(\mu') \ln \frac{\mu^2}{\mu'^2} + B(\mu') \right],
\end{equation}
namely, these coefficients should be evaluated in the necessary approximation with respect to the coupling constant and the coefficients of terms of the series are different for various processes. The first terms of the following expansions:
\begin{equation}
A(\mu') = \sum\limits_{i=0}^{\infty} A^{(n)} \left( \frac{\alpha_s^{} (\mu')}{\pi} \right)^n, \hspace{0.5cm} B(\mu') = \sum\limits_{i=0}^{\infty} B^{(n)} \left( \frac{\alpha_s^{} (\mu')}{\pi} \right)^n
\end{equation}
refer to leading logarithmic (LL) approximation and leading order with respect to the $\alpha_s^{}$ (LO):~\cite{Sun:2012vc}
\begin{equation}
A^{(1)} = C_A^{}, \hspace{0.5cm} B^{(1)} = - \frac{11 C_A^{} - 2 N_f^{}}{6} - \frac{C_A^{}}{2} \delta_{c8}
\end{equation}
the quantity $\delta_{c8}$ is equal to zero in case of the color singlet state of the produced \mbox{$c\bar{c}$-pair} and is equal to one if the final color state is octet, $C_A^{} = N_c^{}$ is an eigenvalue of Casimir operator of the adjoint representation of the SU(3) group, $N_f$ is a number of quark flavors. As the NRQCD-based considerations demonstrate, in gluon-gluon fusion and quark-antiquark annihilation partonic subprocesses, the final state of $c\bar{c}$-pair with the subsequent hadronization into $J/\psi$ can be produced only in the color octet state, therefore an expression for the perturbative Sudakov factor in LL-LO approximation reads
\begin{equation}
S_P^{} (\mu, \mu_b, b_T^{}) = \frac{C_A^{}}{\pi} \int\limits_{\mu_b^2}^{\mu^{2}} \frac{d\mu'^{2}}{\mu'^{2}} \alpha_s (\mu') \left[ \ln \frac{\mu^{2}}{\mu'^2} - \left( \frac{11-2N_f/C_A}{6} + \frac{1}{2} \right) \right] + \mathcal{O}(\alpha_s).
\end{equation}
In addition, the presription~\cite{Boer:2014tka} for the initial scale $\mu_b'^{} = b_0^{}/(b_T^{} + b_0^{}/\mu)$ and cut-off~\cite{Collins:1984kg} for the impact parameter $b_T^{*} (b_T^{}) = b_T^{}/\sqrt{1+(b_T^{}/b_{T,\,\text{max}}^{})^2}$ should be done to stay in perturbative regime $b_0^{}/\mu \leqslant b_T^{} \leqslant b_{T, \text{max}}$ during the TMD PDF evolution and for the Sudakov factor $S_P^{}$ to be valid and of the same sign. We use the common value $b_{T, \text{max}} = 1.5$ GeV$^{-1}$ for value of cut-off parameter. Some other forms of prescription and values of cut-off limit can be also used in the phenomenological studies.~\cite{Bor:2022fga} Moreover, the suppression of the large values of $b_T^{}$ is implemented with the nonperturbative Sudakov factor $S_{NP}^{}$, we use the following parameterization for it:~\cite{Aybat:2011zv}
\begin{equation}
\tilde{S}_{NP}^{} (x, b_T^{}, \mu_{}^{}) = \frac{1}{2} \left[ g_1^{} \ln \frac{\mu_{}^{}}{2Q_{NP}^{}} + g_2^{} \left( 1 + 2 g_3^{} \ln \frac{10 x x_0^{}}{x_0^{} + x} \right) \right] b_T^2
\end{equation}
with these values of the parameters: $g_1^{} = 0.184$ GeV$^2$, $g_2^{} = 0.201$ GeV$^2$, $g_3^{} = -0.129$, $x_0^{} = 0.009$, $Q_{NP}^{} = 1.6$ GeV. It was extracted for quarks as initial partons from the SIDIS experimental data. Because there's no such direct opportunity to obtain it for gluons as initial partons, the nonperturbative Sudakov factor $S_{NP}^{}$ is usually rescaled with the color factor $C_A^{} / C_F^{}$, where $C_F^{} = (N_c^2 - 1)/(2N_c^{})$, and further applied to gluons.~\cite{Balazs:2007hr} The factor $S_{NP}^{}$ is a main source of the nonperturbative input for the TMD PDFs within the SGR approach. The cross section includes it in the exponentiated form $e^{-S_{NP}^{}}$ which implements a Gaussian shape for the $b_T^{}$-distribution and, if the inverse Fourier transform is done, for the $q_T^{}$-distribution. The expression for cross section includes two TMD PDFs, that's why the final Sudakov factor is taken as a sum of two terms for each parton:
\begin{equation}
S_{NP}^{} (x_1^{}, x_2^{}, b_T^{}, \mu_{}^{}) = \tilde{S}_{NP}^{} (x_1^{}, b_T^{}, \mu_{}^{}) + \tilde{S}_{NP}^{} (x_2^{}, b_T^{}, \mu_{}^{}).
\end{equation}

Within the perturbative regime, the TMD PDFs are determined by their so-called perturbative tails, in the leading approximation they are just reduced to the collinear distributions as follows:~\cite{Sun:2011iw}
\begin{equation}
\hat{f}_1^{}(x, b_T^{}, \mu'^{}_{b}) = f (x, \mu'^{}_{b}) + \mathcal{O} (\alpha_s) + \mathcal{O} (b_T^{} \Lambda_{\text{QCD}}).
\end{equation}

Gathering all the details for calculation of $c\bar{c}$-pair production cross section in the SGR approach together, we obtain the following expression for differential cross section:
\begin{align}
\frac{d\sigma^{TMD}}{dp_{T}^{c\bar{c}} \, dy_{c\bar{c}}^{} \, dM_{c\bar{c}}^2} = \frac{p_T^{c\bar{c}}}{s} \int  db_T^{} \, b_T^{} \, &J_0(b_T^{}p_T^{c\bar{c}} ) \times \nonumber\\
& \times e^{-S_P^{}} \, e^{-S_{NP}^{}} \, f (x_1^{}, \mu'^{}_{b*}) \,f (x_2^{}, \mu'^{}_{b*}) \, \hat{\sigma}(M_{c\bar{c}}^2),
\label{eq:tmd_cs}
\end{align}
with a transverse momentum $p_T^{c\bar{c}}$ of the $c\bar{c}$-pair, its rapidity  $y_{c\bar{c}}^{}$, invariant mass $M_{c\bar{c}}^{}$, the first kind Bessel function of the zeroth order $J_0^{}$ and total cross sections of $c\bar{c}$-pair production in $gg \rightarrow c\bar{c}$ partonic subprocess
\begin{equation}
\hat{\sigma} (\hat{s}) = \frac{\pi\alpha_s^2}{3\hat{s}} \left[ \left(1 + w + \frac{w^2}{16} \right) \ln \left( \frac{1+\sqrt{1-w}}{1-\sqrt{1-w}} \right) - \left( \frac{7}{4}  + \frac{31 w}{16}\right) \sqrt{1-w} \right]
\end{equation}
and in $q\bar{q} \rightarrow c\bar{c}$ partonic subprocess
\begin{equation}
\hat{\sigma} (\hat{s}) = \frac{4 \pi \alpha_s^2}{27 \hat{s}} \left( w + 2 \right) \sqrt{1-w},
\end{equation}
where $w = 4m_c^2/\hat{s}$ and $\hat{s} = (q_1^{} + q_2^{})^2 = (p^c + p^{\bar{c}})^2$ with the mass $m_c^{}$ of a charm quark.

\section{Collinear factorization}

For large values of transverse momentum $p_T^{} \gg M_{J/\psi}^{}$ of charmonium another approach for factorization of cross section should be used, it is the Collinear Parton Model (CPM) or Collinear factorization. The on-shell four-momenta of the initial partons within the CPM are determined as $q_{1,2}^{\mu} = x_{1,2}^{} p_{1,2}^{\mu}$ with momenta of the incoming protons $p_{1,2}^{\mu}$. The corresponding factorization theorem allows to write down cross section expression in the standard form of convolution of collinear PDFs and hard-scale partonic cross section:
\begin{equation}
d\sigma^{CPM} = \int dx_1^{} \int  dx_2^{} \, f (x_1^{}, \mu) f (x_2^{}, \mu) \,d\hat{\sigma},
\label{eq:cpm1}
\end{equation}
where $x_{i}^{}$ are also parton momentum fraction, $f(x, \mu)$ are collinear PDFs for gluons, quarks or antiquarks, $d\hat{\sigma}$ is a hard partonic cross section which reads
\begin{equation}
d\hat{\sigma} = (2\pi)^4 \delta^{(4)} (q_1^{} + q_2^{} - p_{c}^{} - p_{\bar{c}}^{} - k^{}) \frac{\overline{|\mathcal{M}(2 \rightarrow 3)|^2}}{I} \frac{d^3 p_{c}^{}}{(2\pi)^3 2 p_{c0}^{}} \frac{d^3 p_{\bar{c}}^{}}{(2\pi)^3 2 p_{\bar{c}0}^{}} \frac{d^3 k}{(2\pi)^3 2 k_{0}^{}}
\label{eq:cpm2}
\end{equation}
with $I = 2 x_1^{} x_2^{} s$ as a flux factor, $\overline{|\mathcal{M}(2 \rightarrow 3)|^2}$ as a squared amplitude of the partonic subprocesses $gg \rightarrow c\bar{c} g$ and $q\bar{q} \rightarrow c\bar{c} g$ averaged over initial spin and color degrees of freedom. The corresponding amplitudes were calculated using the FeynArts~\cite{Hahn:2000kx} and FeynCalc~\cite{Shtabovenko:2020gxv} packages.

We should integrate in the formula~(\ref{eq:cpm2}) over variables which are independent of $M_{c\bar{c}}^{}$, $p_T^{c\bar{c}}$ and $y_{c\bar{c}}$. For this purpose, we may write squared amplitudes in the $c\bar{c}$-pair center-of-mass frame in which expressions of the final quarks momenta read
\begin{equation}
p_{c/\bar{c}}^{\mu} = \left( \frac{M_{c\bar{c}}}{2}, \pm w_0^{} \sin\theta \sin\varphi, \pm w_0^{} \sin\theta \cos\varphi, \pm w_0^{} \cos\theta \right),
\end{equation}
\begin{equation}
w_0^{} = \frac{M_{c\bar{c}}}{2} \sqrt{1 - \frac{4m_c^2}{M_{c\bar{c}}^2}}
\end{equation}
with a polar angle $\theta$ and an azymuthal angle $\varphi$ in the $c\bar{c}$-pair rest frame. The momenta of initial partons should evidently be written in the same frame which can done by the Lorentz transformation with the velocity $\vec{\upsilon}$ of the $c\bar{c}$-pair rest frame in the laboratory frame:
\begin{equation}
\vec{\upsilon} = \upsilon \frac{\vec{p}^{\,c\bar{c}}}{|\vec{p}^{\,c\bar{c}}\,|} = \frac{(\vec{p}_T^{\,c\bar{c}}, M_T^{} \sinh y_{c\bar{c}}^{})}{M_T^{} \cosh y_{c\bar{c}}^{}},
\end{equation}
where $M_T^{} = \sqrt{M_{c\bar{c}}^2 + (p_T^{c\bar{c}})^2}$ is a transverse mass of the final $c\bar{c}$-pair. Thus, if we substitute equation~(\ref{eq:cpm2}) to (\ref{eq:cpm1}) and evaluate all analytical integrations, we obtain the expression for differential cross section of $c\bar{c}$-pair production in the collinear factorization as follows:
\begin{align}
\frac{d\sigma^{CPM}}{dp_{T}^{c\bar{c}} \, dy_{c\bar{c}} \, dM_{c\bar{c}}^2} &= \frac{p_{T}^{c\bar{c}}}{32 (2\pi)^4 s} \sqrt{1- \frac{4m_c^2}{M_{c\bar{c}}^2}} \times \nonumber\\
& \times \int\limits_{x_1^{\text{min}}}^1 \frac{dx_1^{} f (x_1^{}, \mu) f (x_2^{}, \mu)}{x_1^{} x_2^{} (s x_1^{} - M_T^{} \sqrt{s} e^{y_{c\bar{c}}^{}})} \int\limits_{\Omega} \overline{|\mathcal{M}|^2} \sin\theta\, d\theta \,d\varphi,
\end{align}
where the momentum fraction $x_2^{}$ and the lower integration limit $x_{1}^\text{min}$ are
\begin{equation}
x_2^{} = \frac{x_1^{} M_T^{} \sqrt{s} e^{-y_{c\bar{c}}^{}}  - M_{c\bar{c}}^2}{s x_1^{} - M_T^{}\sqrt{s}e^{y_{c\bar{c}}^{}}},~~~ x_1^{\text{min}} = \frac{M_T^{} \sqrt{s} e^{y_{c\bar{c}}^{}}  - M_{c\bar{c}}^2}{s - M_T^{} \sqrt{s} e^{-y_{c\bar{c}}^{}}}.
\end{equation}

In order to correspond to the SGR calculations and perform the proper matching of TMD and CPM results, we make calculation in LO with respect to $\alpha_s^{}$ and take into account both gluon-gluon fusion and quark-antiquark annihilation subprocesses.

\section{Improved Color Evaporation model}

We work with Improved Color Evaporation model~\cite{Ma:2016exq} as an approach to describe hadronization of produced quarks into charmonium state. Within the ICEM, the $c\bar{c}$-pair is considered to be produced in a partonic subprocess with the invariant mass $M_{c\bar{c}}^{}$, then the series of gluon emissions and exchanges with other color sources of the hard subprocess happens --- so the bound state of charmed quarks, charmonium with mass $M_{\mathcal{C}}^{}$, is formed this way. Speaking briefly, this model is based on the traditional Color Evaporation model~\cite{Fritzsch:1977ay, Halzen:1977rs} with two key technical modifications.

Firstly, the momenta of produced quark-pair and charmonium are related by the following ratio of proportionality: $p_{\mu}^{J/\psi} =  (M_{J/\psi}^{}/M_{c\bar{c}}^{})\, p_{\mu}^{c\bar{c}}$, which represents the momentum shift effect during the hadronization. To mention the second feature, we will move on to expression of $J/\psi$ production cross section:
\begin{equation}
\frac{d\sigma}{d\vec{p}^{\,\, J/\psi}} = F^{J/\psi} \int\limits_{M_{J/\psi}^{2}}^{4M_{D}^{2}} dM_{c\bar{c}}^{2} \int d^3 p^{c\bar{c}} \delta^{(3)} \left( \vec{p}^{\,\,J/\psi} - \frac{M_{J/\psi}^{}}{M_{c\bar{c}^{}}} \vec{p}^{\,\,c\bar{c}} \right) \frac{d\sigma}{d\vec{p}^{\, c\bar{c}}\,dM_{c\bar{c}}^{2} }.
\end{equation}
But because we study transverse momenta spectra, we'll write the relation for differential cross section over $p_T^{J/\psi}$ as follows
%\begin{align}
%\frac{d\sigma}{dp_{T}^{\,\,J/\psi} dy} = F^{J/\psi} \int\limits_{M_{J/\psi}^{2}}^{4M_{D}^{2}} dM_{c\bar{c}}^{2} \int d^2 p_{T}^{c\bar{c}} \, \delta^{(2)} &\left( \vec{p}_{T}^{\,\,J/\psi} - \frac{M_{J/\psi}^{}}{M_{c\bar{c}^{}}} \vec{p}_{T}^{\,\,c\bar{c}} \right) \times \nonumber \\ 
%&\times  \frac{M_{c\bar{c}}^{}}{M_{J/\psi}^{}} \, \frac{d\sigma}{dp_{T}^{\,c\bar{c}} \, dy \, dM_{c\bar{c}}^2}.
%\label{eq:icem}
%\end{align}
\begin{align}
\frac{d\sigma}{dp_{T}^{\,\,J/\psi} dy} = F^{J/\psi} \int\limits_{M_{J/\psi}^{2}}^{4M_{D}^{2}} dM_{c\bar{c}}^{2} \, \frac{M_{c\bar{c}}^{}}{M_{J/\psi}^{}} \, \left( \frac{d\sigma}{dp_{T}^{\,c\bar{c}} \, dy \, dM_{c\bar{c}}^2} \right) \Bigg|_{p_T^{c\bar{c}}=\frac{M_{c\bar{c}}^{}}{M_{J/\psi}^{}} p_T^{J/\psi}}
\label{eq:icem}
\end{align}
where, instead of integration from $(2m_c^{})^2$ in the CEM, the ICEM prescribes the lower limit to be $M_{J/\psi}^2$. The upper limit corresponds to threshold of hidden-charm meson production. The subscript of rapidity $y$ is omitted here, as it is defined with the ratio of momentum components, that is why it doesn't change when the momentum shift is taken into account and the rapidity is actually the same for both $c\bar{c}$-pair and $J/\psi$-meson. We use the following values for ICEM calculations: charm quark mass $m_c^{} = 1.2$ GeV, $D$-meson mass $M_D^{} = 1.84$ GeV and $J/\psi$-meson mass $M_{J/\psi} = 3.096$~GeV. The cross section is also multiplied by the phenomenological factor $F^{J/\psi}$ that can be interpreted as a probability of hadronization of $c\bar{c}$-pair into the $J/\psi$ state of charmonium.

The final expressions, ready for numerical evaluation of cross section within the ICEM, are obtained by substitutions: equation~(\ref{eq:tmd_cs}) is inserted into formula~(\ref{eq:icem}) which gives us the following result for the SGR approach:
\begin{align}
&\frac{d\sigma^{TMD}}{dp_{T}^{J/\psi} \, dy} = F^{J/\psi} \,\frac{p_T^{\,J/\psi}}{s} \int\limits_{M_{J/\psi}^{2}}^{4M_{D}^{2}} dM_{c\bar{c}}^{2} \, \frac{M^2_{c\bar{c}}}{M^2_{J/\psi}} \times \nonumber\\
&\times \int\limits_0^{\infty} db_T^{} \, b_T^{} \, J_0 \left( \frac{M_{c\bar{c}}}{M_{J/\psi}} p_T^{J/\psi}  b_T^{} \right) e^{-S_P^{}} \, e^{-S_{NP}^{}} \, f (x_1^{}, \mu'^{}_{b*}) \,f (x_2^{}, \mu'^{}_{b*}) \, \hat{\sigma}(M_{c\bar{c}}^2),
\end{align}
and equation (\ref{eq:cpm2}) is inserted into formula (\ref{eq:icem}) for calculation in the CPM which provides us with the following expression:
\begin{align}
\frac{d\sigma^{CPM}}{dp_{T}^{J/\psi} \, dy} = F^{J/\psi} \,\frac{p_{T}^{J/\psi}}{32 (2\pi)^4 s}& \int\limits_{M_{J/\psi}^{2}}^{4M_{D}^{2}} dM_{c\bar{c}}^{2} \, \frac{M^2_{c\bar{c}}}{M^2_{J/\psi}} \sqrt{1- \frac{4m_c^2}{M_{c\bar{c}}^2}}  \times \nonumber\\
&\times \int\limits_{x_1^{\text{min}}}^1 \frac{dx_1^{} f (x_1^{}, \mu) f (x_2^{}, \mu)}{x_1^{} x_2^{} (s x_1^{} - M_T^{} \sqrt{s} e^{y} )} \int\limits_{\Omega} \overline{|\mathcal{M}|^2} \sin\theta \,d\theta\, d\varphi,
\end{align}
where the momentum shift substitution should be properly made.

\section{Matching scheme}

The region of intermediate transverse momentum $p_T^{} \sim M$ is out of domains of applicability of TMD and CPM factorization theorems. Instead, some matching approaches is often used to describe this kinematic region. Traditional one is the Collins-Soper-Sterman (CSS) scheme,~\cite{Collins:1981uk, Collins:2016hqq} in which the TMD-term and fixed-order CPM-term are summed and then the double-counted contributions are subtracted. Another scheme for matching, and the scheme we use in the study, is the Inverse-Error Weighting~\cite{Echevarria:2018qyi} (InEW). Within the InEW, the weighted TMD and CPM-terms are added up, so that no double-counted terms appear:
\begin{equation}
\overline{d\sigma}(p_T^{},Q) = \omega_1^{} \, d\sigma^{TMD}(p_T^{},Q) +
\omega_2^{} \, d\sigma^{CPM}(p_T^{},Q).
\end{equation}
The weights for both terms are determined with power corrections of the corresponding factorization theorems $\Delta_{TMD}^{}$ and $\Delta_{CPM}^{}$, they are interpreted as accuracies of the TMD and CPM approximations at the given value of transverse momentum $p_T^{}$ of the final state:
\begin{equation}
\omega_1^{} = \frac{\Delta^{-2}_{{TMD}}}{\Delta^{-2}_{{TMD}} + \Delta^{-2}_{{CPM}}},~~~~~~~ \omega_2^{} = \frac{\Delta^{-2}_{{CPM}}}{\Delta^{-2}_{{TMD}} + \Delta^{-2}_{{CPM}}},
\end{equation}
\begin{equation}
\Delta_{{TMD}}^{} = \left( \frac{p_T^{}}{Q} \right)^2 + \left( \frac{m}{Q} \right)^2,~~~~~~~ \Delta_{{CPM}}^{} = \left( \frac{m}{p_T^{}} \right)^2 \cdot \left( 1 + \ln^2 \left( \frac{Q_T^{}}{p_T^{}}  \right)   \right),
\end{equation}
the value $m$ is a scale of the hadronic mass order which is about $1$ GeV, $Q_T^{} = \sqrt{Q^2 + p_T^2}$ and $Q = M_{c\bar{c}}^{}$ in our study. At the small transverse momenta, the TMD weight $\omega_1^{}$ is close to one, and it dominates up to $p_T^{} \sim Q$. At high values of transverse momentum, the CPM weight $\omega_2^{}$ prevails and, at the intermediate region $p_T^{} \sim Q$, its dominance vanishes and the matched cross section $\overline{d\sigma}$ is determined by terms of both factorizations. 

This scheme, being based on the inverse-variance weighting approach for aggregating several random variables, provides us with the expression for evaluating of uncertainty of the matching procedure in terms of uncertainties of factorization theorems:
\begin{equation}
\Delta \overline{d\sigma} = \frac{ \Delta_{{TMD}}^{}
\Delta_{{CPM}}^{}}{\sqrt{\Delta_{{TMD}}^{2} + \Delta_{{CPM}}^{2}}}
\overline{d\sigma}.
\label{eq:uncertainty}
\end{equation}
The magnitude of uncertainty is naturally minimized in the domains of applicability of factorization theorems and is maximized at the intermediate region of transverse momentum, where no factorization theorem fully determines the cross section value.

\section{Results of calculation}

Numerical calculations made in our study were performed using the library for numerical integration CUBA,~\cite{Hahn:2004fe} a maximum relative error of calculation was $1\%$. The collinear PDFs used for the fixed-order calculations and for modelling of TMD PDFs within the SGR were the tabulated MSTW2008LO distributions~\cite{Martin:2009iq} as they were probed for the SGR PDFs evolution in the Refs.~\refcite{Boer:2014tka,Bor:2022fga}. Though, the more up-to-date distributions are evidently the preferred option, the discrepancy between the distribution used in this work and the NNPDF2.3LO~\cite{Ball:2012cx} distribution, which we compared with in one of the previous calculations, was minor and entirely absorbed by the uncertainty due to hard scale variation. We used the invariant mass $M_{c\bar{c}}^{}$ for TMD calculations and the transverse mass $M_T^{} = \sqrt{M_{c\bar{c}}^2 + (p_T^{c\bar{c}})^2}$ for the CPM as factorization scale $\mu$ and renormalization scale $\mu_R^{}$. The parameters utilized in the calculations are as follows:~\cite{ParticleDataGroup:2024cfk} branching ratios of $J/\psi$ decay into lepton pairs \mbox{Br$(J/\psi \rightarrow e^{+} e^{-}) = 0.05971$} and \mbox{Br$(J/\psi \rightarrow \mu^+ \mu^-) = 0.05961$}.

\begin{figure}[b!]
\begin{center}
\begin{minipage}[h]{0.49\linewidth}
\includegraphics[width=6.6cm]{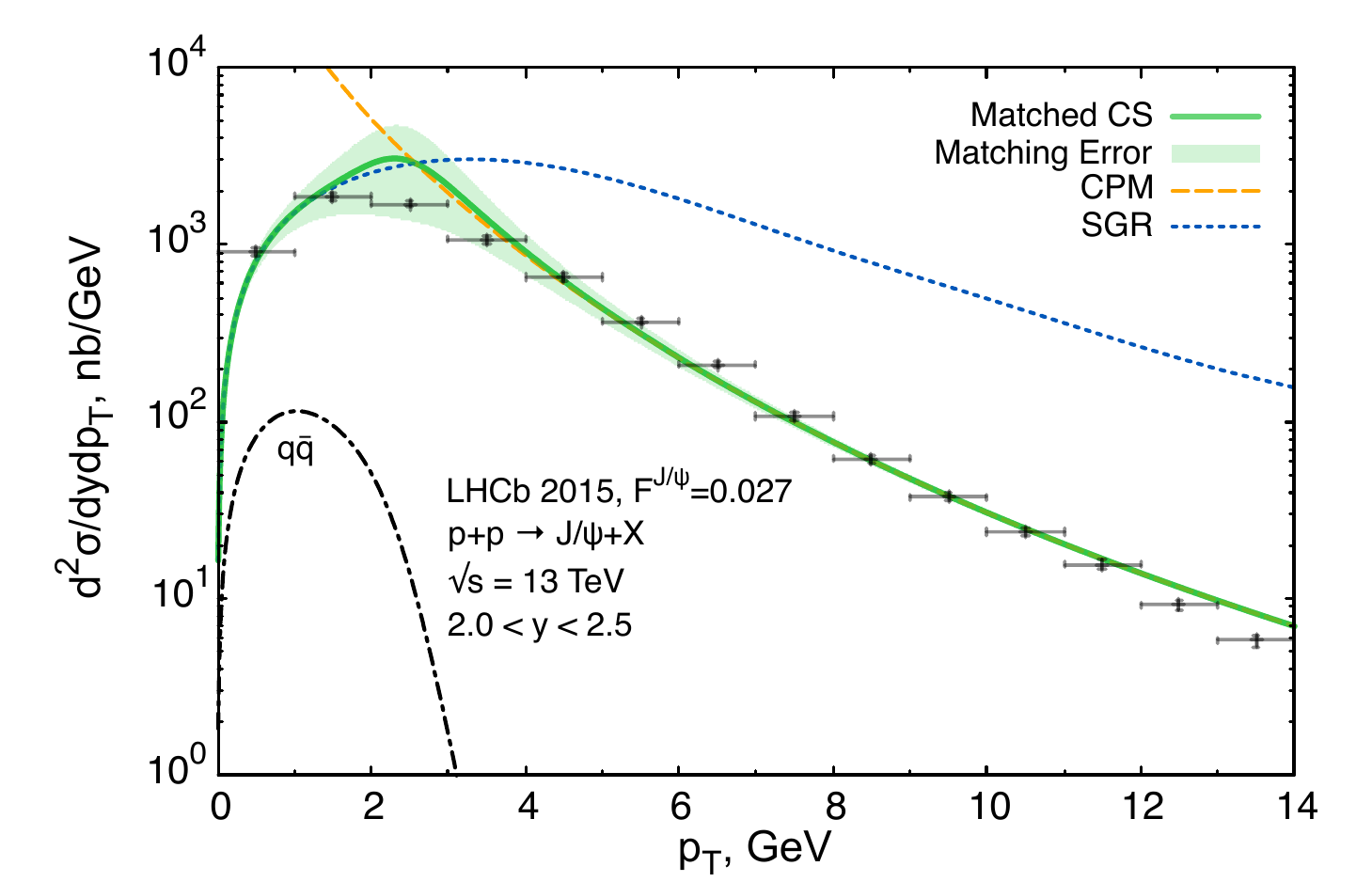}
\end{minipage}
\hfill
\begin{minipage}[h]{0.49\linewidth}
\includegraphics[width=6.6cm]{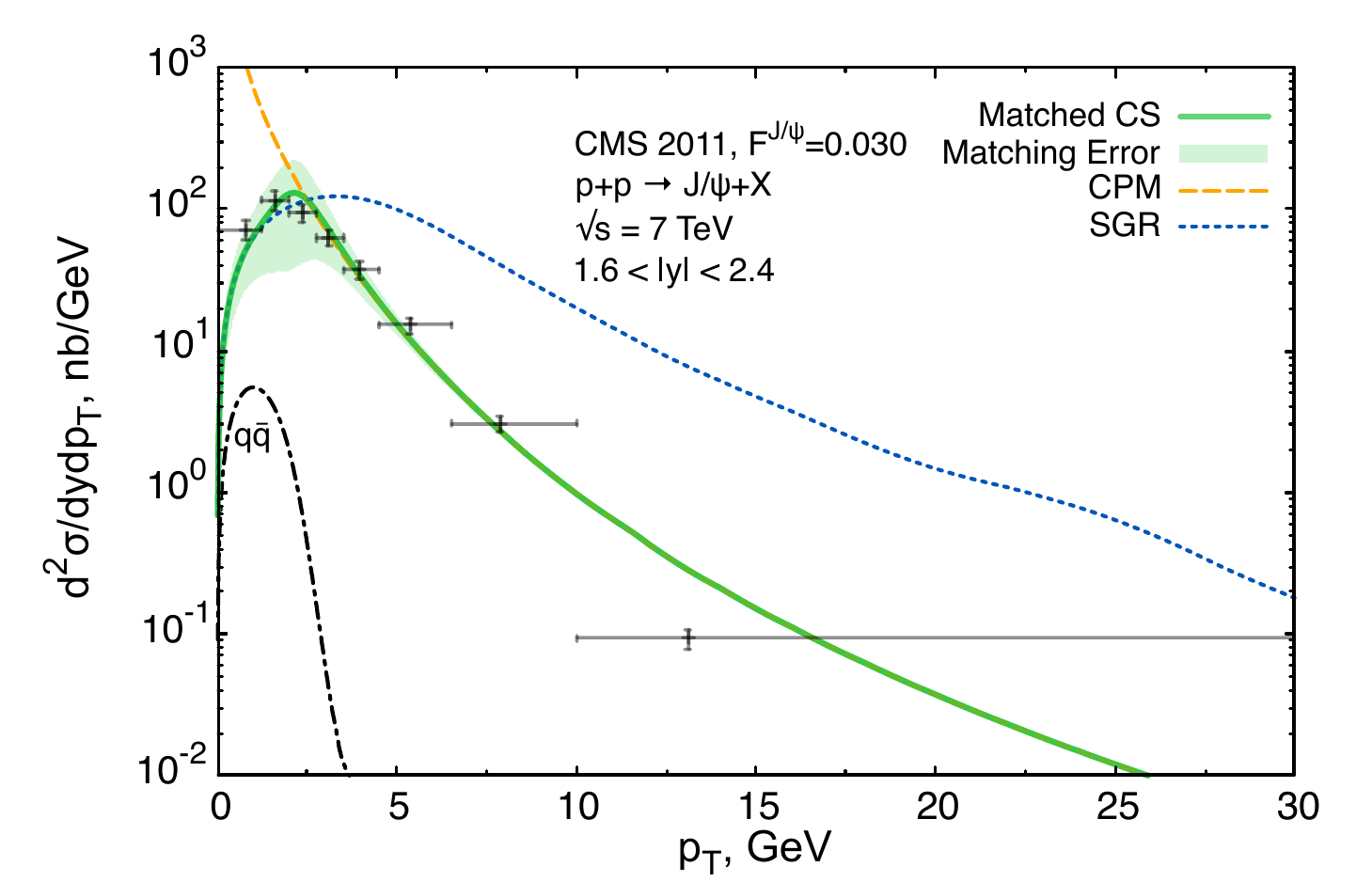}
\end{minipage}
\caption{The cross section of $J/\psi$ production versus transverse momentum at $\sqrt{s} = 13$ TeV (left panel) and $\sqrt{s} = 7$ TeV (right panel). The experimental data were taken from the LHCb~\cite{LHCb:2015foc} and CMS~\cite{CMS:2010nis} Collaborations} 
\label{fig:1}
\end{center}

\end{figure}

\begin{figure}[b!]
\begin{center}
\begin{minipage}[h]{0.49\linewidth}
\includegraphics[width=6.6cm]{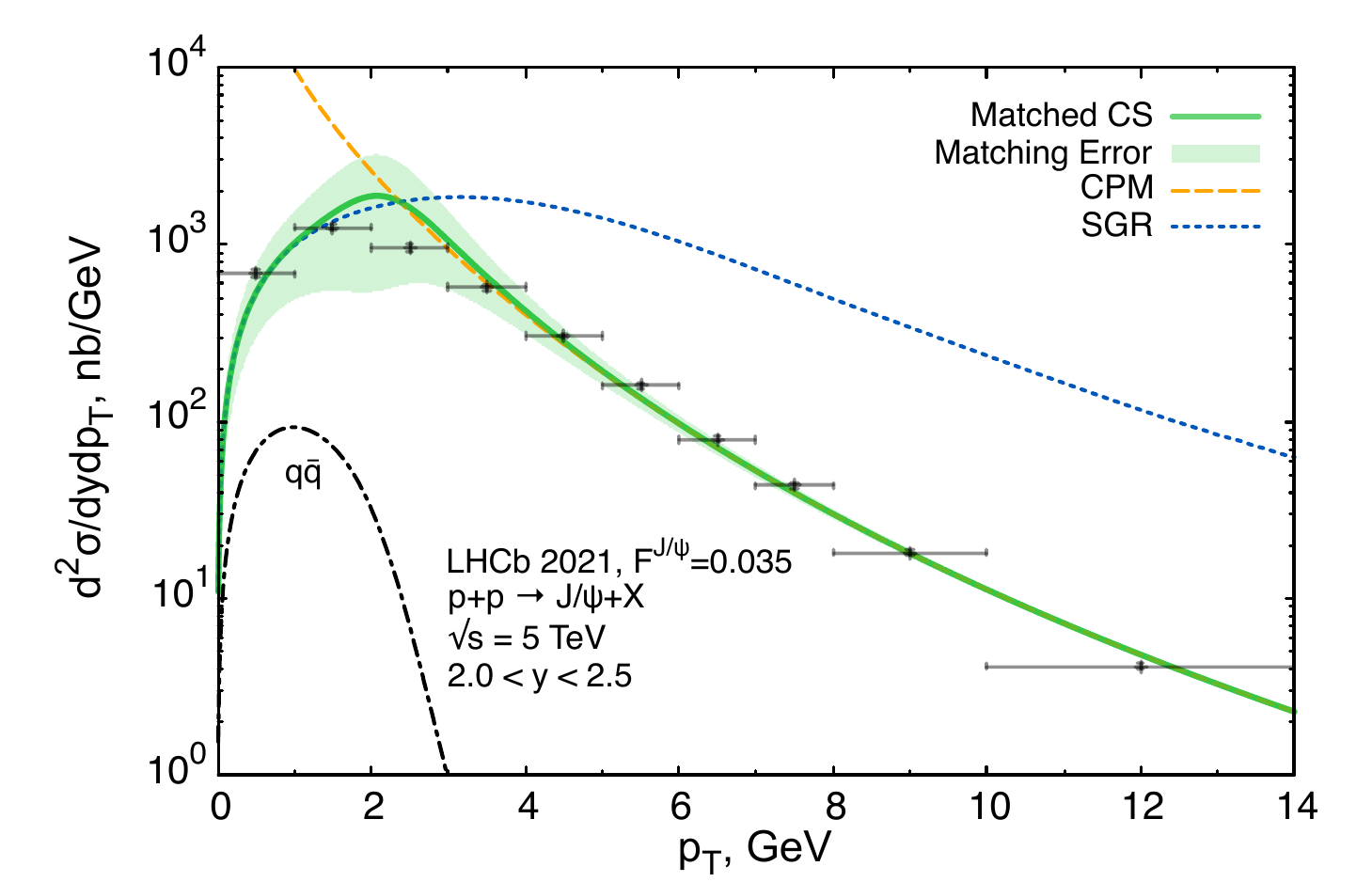}
\end{minipage}
\hfill
\begin{minipage}[h]{0.49\linewidth}
\includegraphics[width=6.6cm]{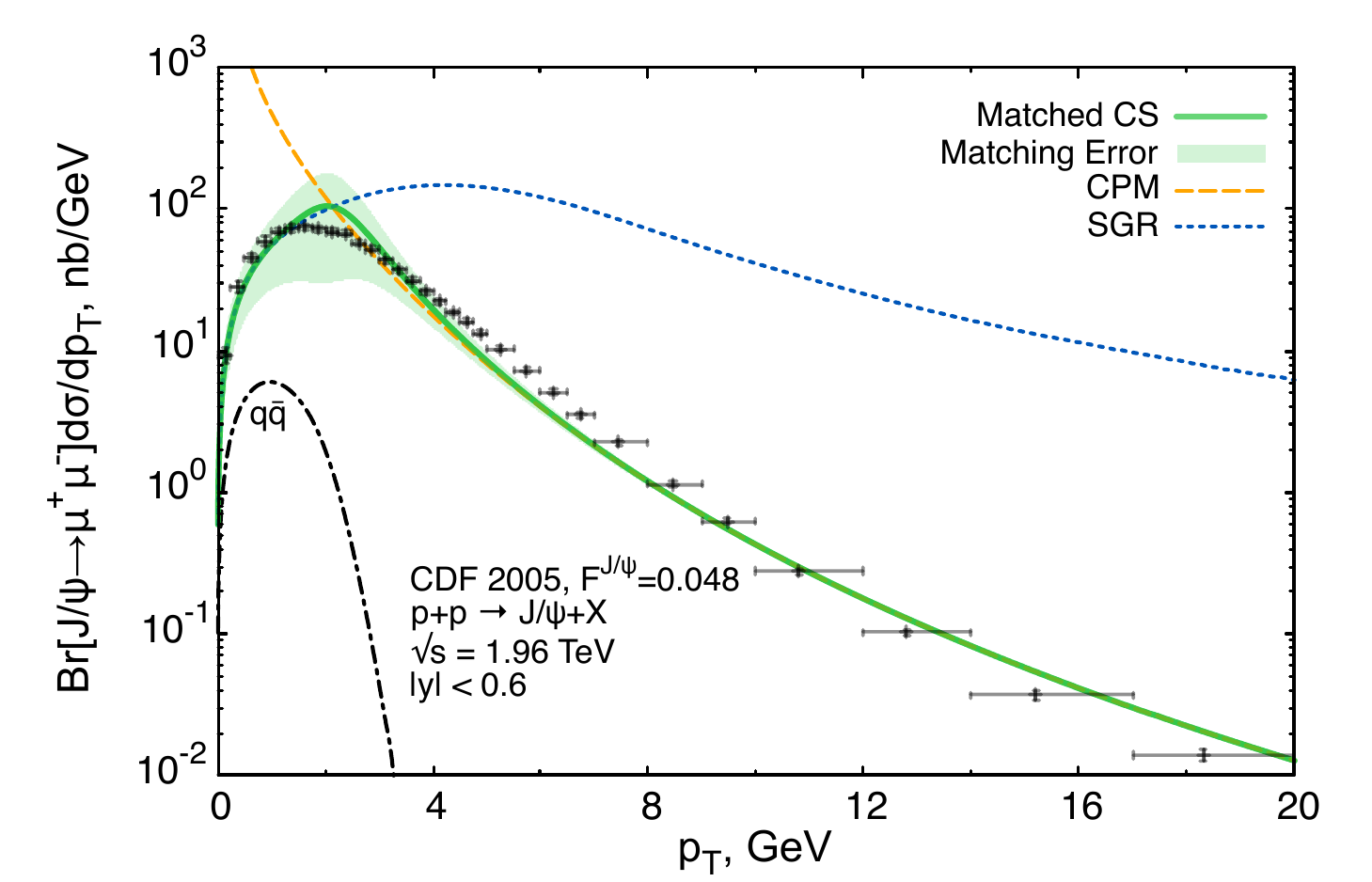}
\end{minipage}
\caption{The cross section of $J/\psi$ production versus transverse momentum at $\sqrt{s} = 5$ TeV (left panel) and $\sqrt{s} = 1.96$ TeV (right panel). The experimental data were taken from the LHCb~\cite{LHCb:2021pyk} and CDF~\cite{CDF:2004jtw} Collaborations} 
\label{fig:2}
\end{center}
\end{figure}

\begin{figure}[b!]
\begin{center}
\begin{minipage}[h]{0.49\linewidth}
\includegraphics[width=6.6cm]{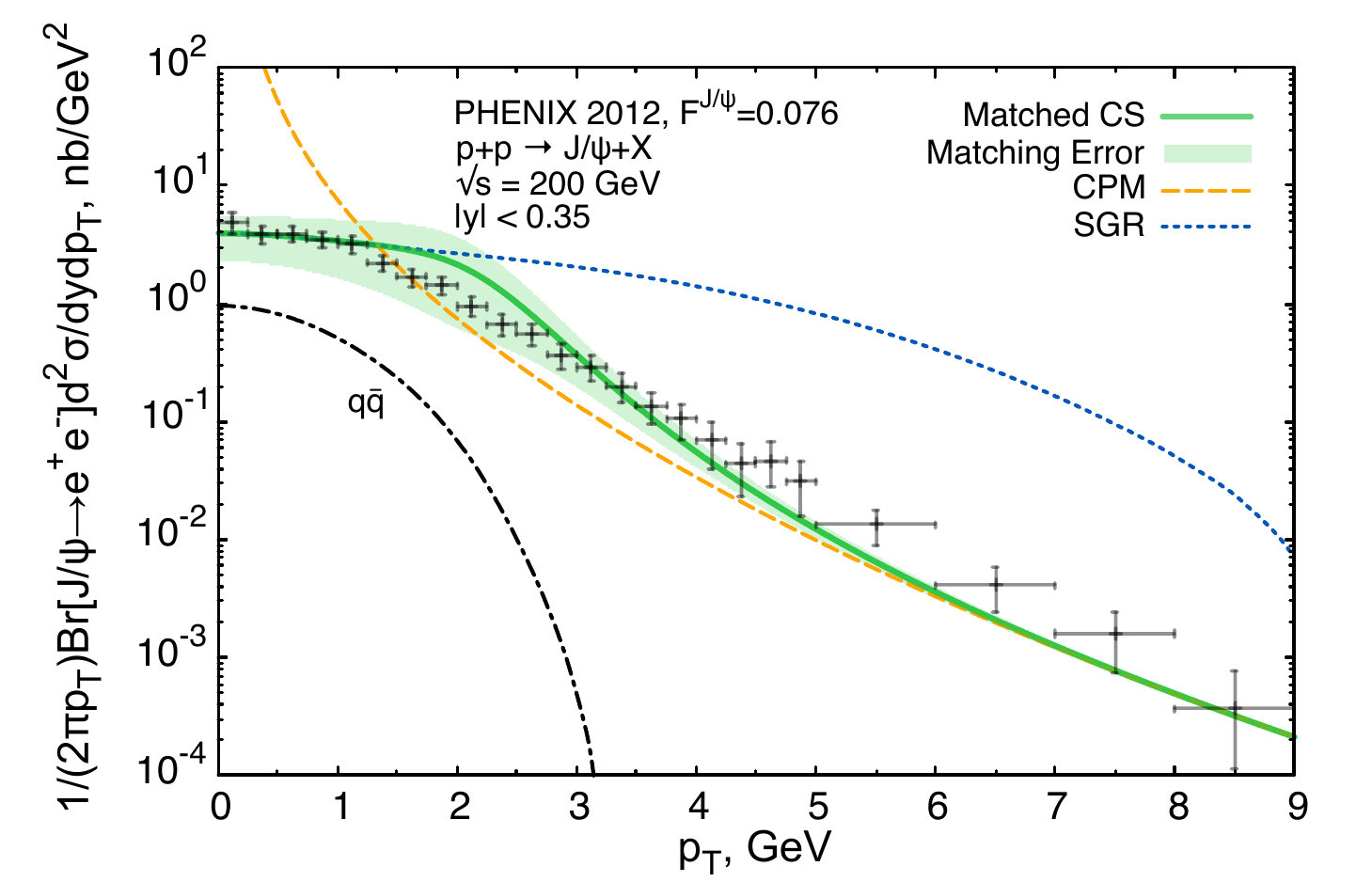}
\end{minipage}
\hfill
\begin{minipage}[h]{0.49\linewidth}
\includegraphics[width=6.6cm]{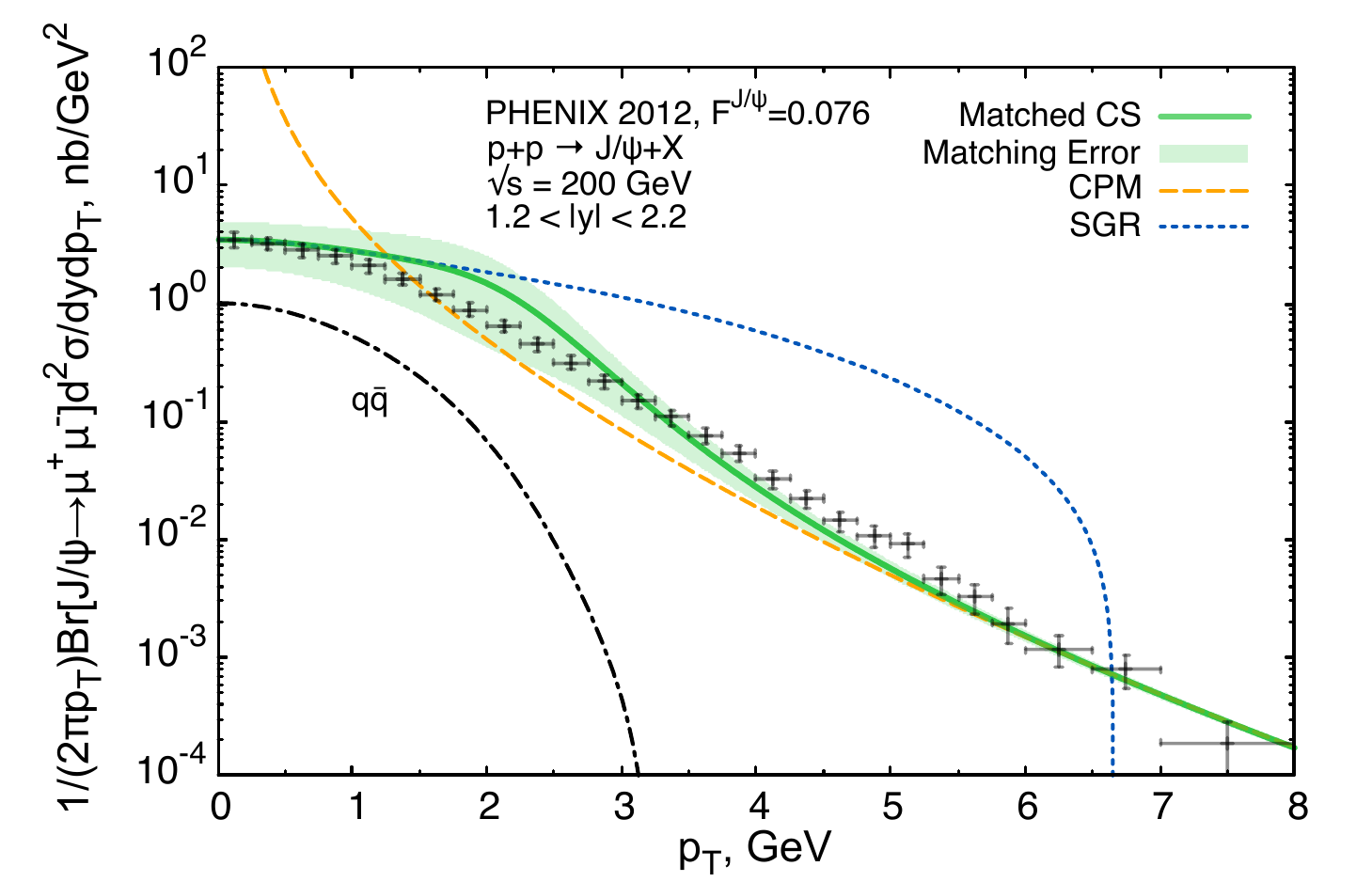}
\end{minipage}
\caption{The cross section of $J/\psi$ production versus transverse momentum at $\sqrt{s} = 200$ GeV. The experimental data were taken from the PHENIX Collaboration~\cite{PHENIX:2011gyb}} 
\label{fig:3}
\end{center}
\end{figure}

\begin{figure}[h!]
\begin{center}
\begin{minipage}[h]{0.49\linewidth}
\includegraphics[width=6.6cm]{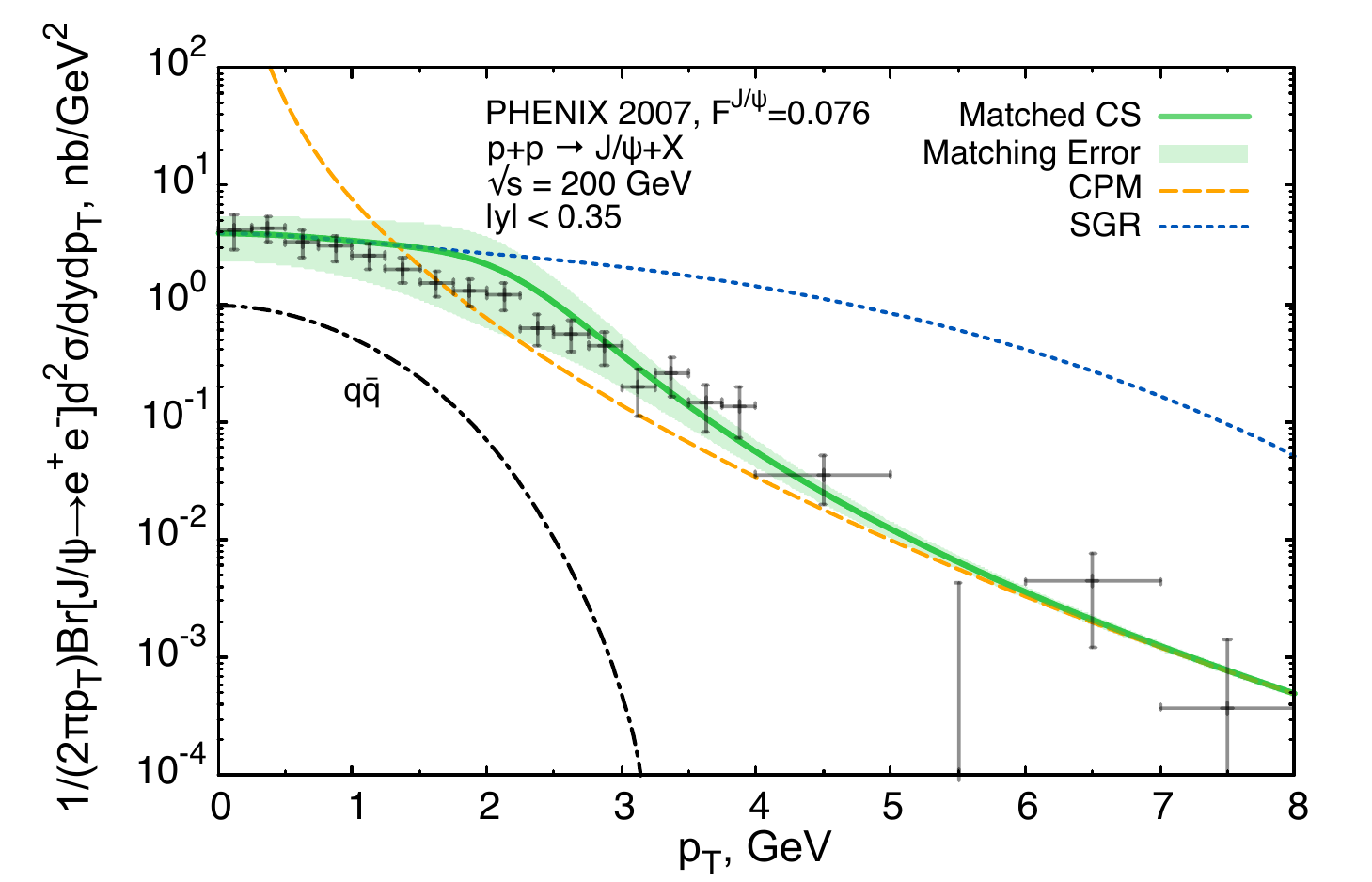}
\end{minipage}
\hfill
\begin{minipage}[h]{0.49\linewidth}
\includegraphics[width=6.6cm]{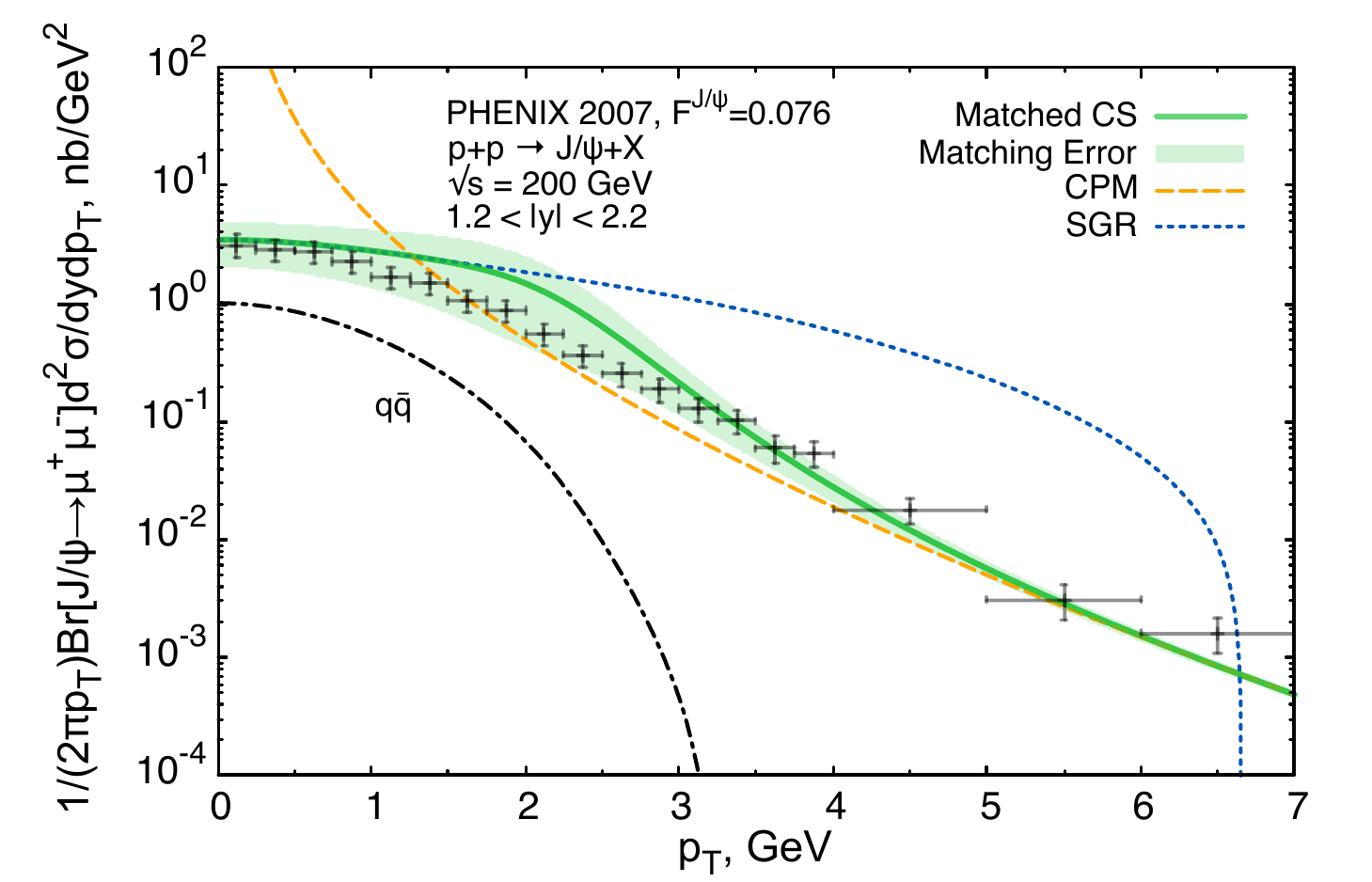}
\end{minipage}
\caption{The cross section of $J/\psi$ production versus transverse momentum at $\sqrt{s} = 200$ GeV. The experimental data were taken from the PHENIX Collaboration~\cite{PHENIX:2011gyb}} 
\label{fig:4}
\end{center}
\end{figure}

\begin{figure}[h!]
\begin{center}
\begin{minipage}[h]{0.49\linewidth}
\includegraphics[width=6.6cm]{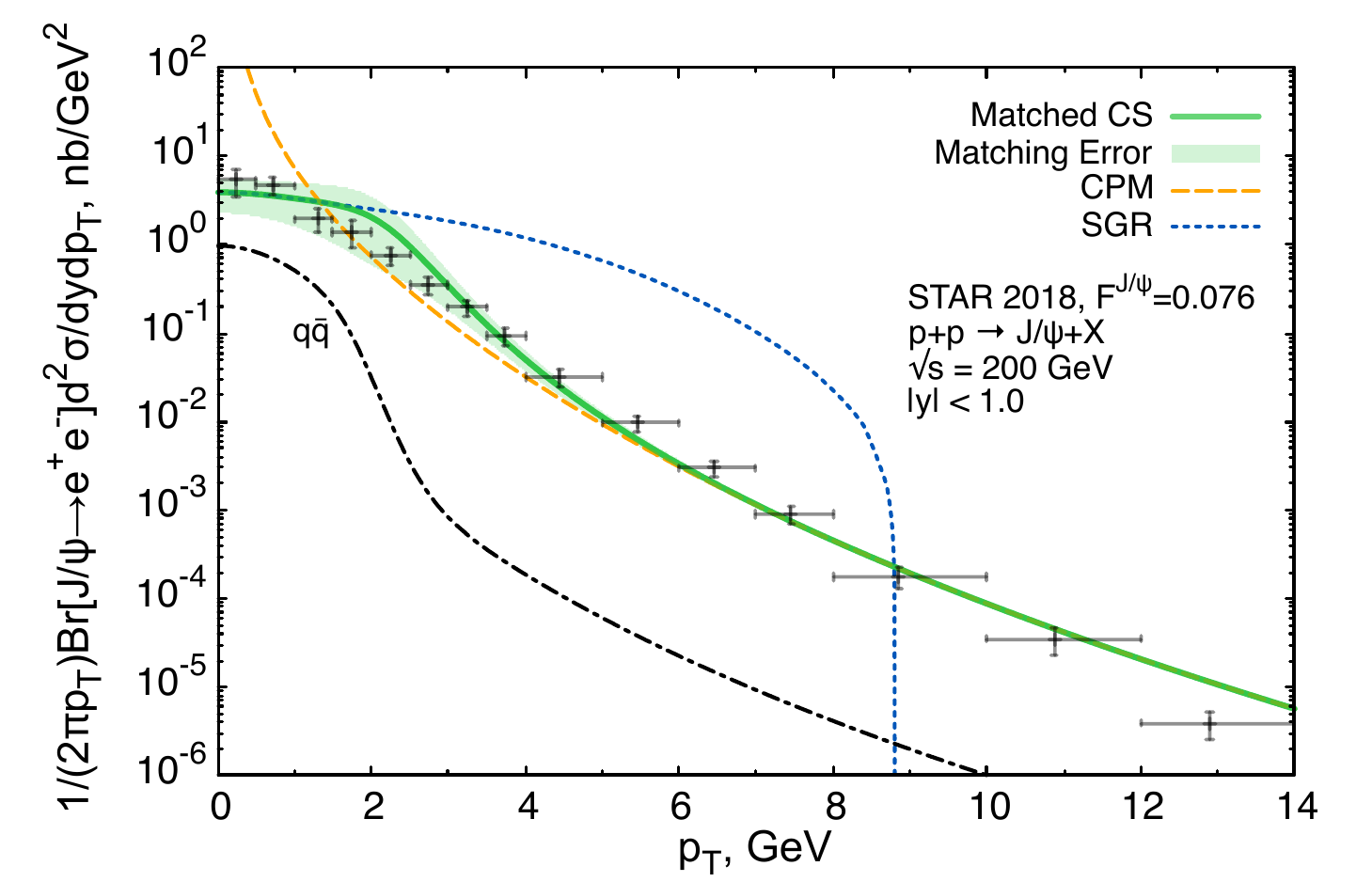}
\end{minipage}
\hfill
\begin{minipage}[h]{0.49\linewidth}
\includegraphics[width=6.6cm]{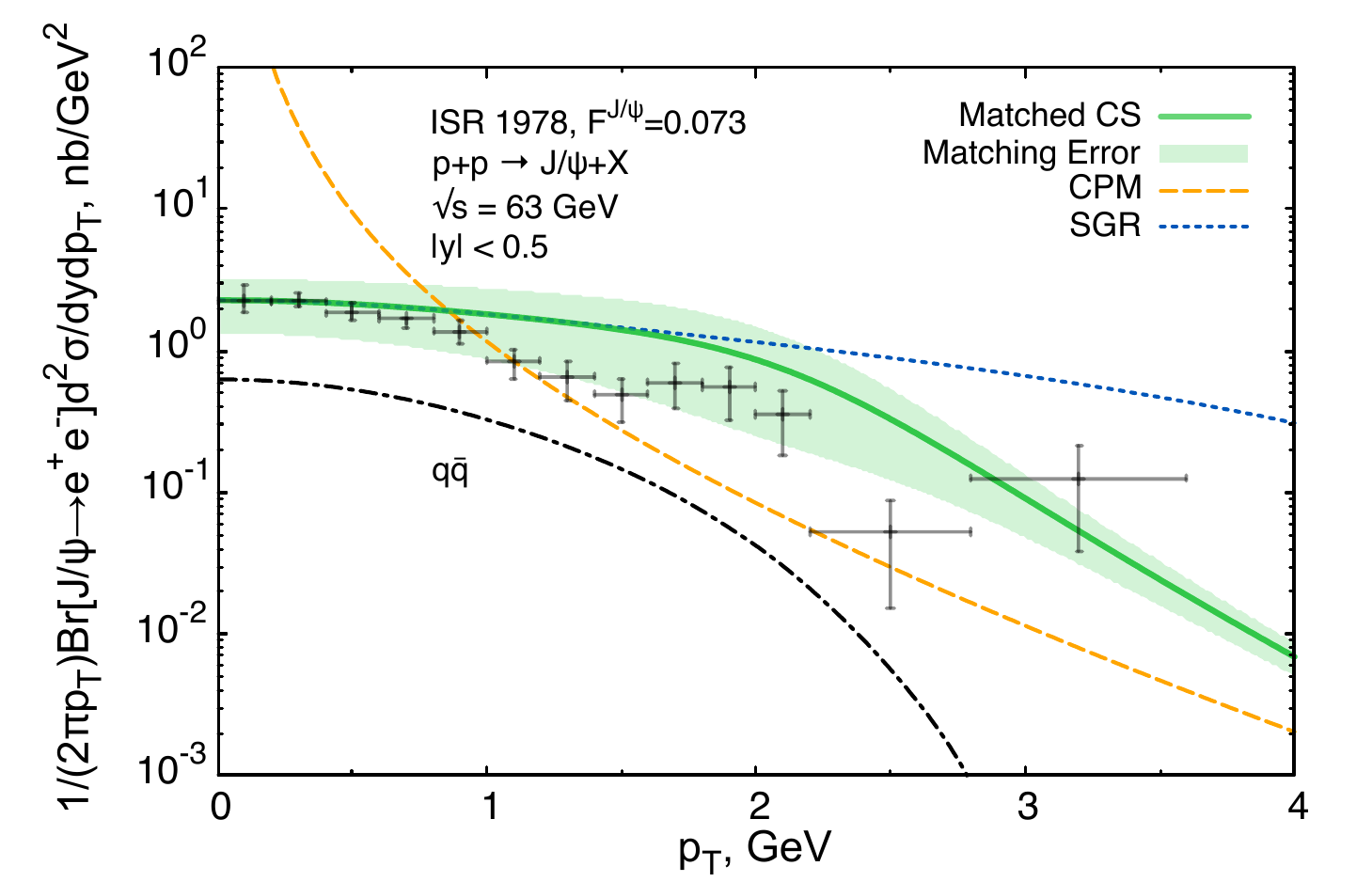}
\end{minipage}
\caption{The cross section of $J/\psi$ production versus transverse momentum at $\sqrt{s} = 200$ GeV (left panel) and $\sqrt{s} = 63$ GeV. The experimental data were taken from the PHENIX~\cite{PHENIX:2011gyb} and ISR~\cite{Clark:1978mg} Collaborations} 
\label{fig:5}
\end{center}
\end{figure}

\begin{figure}[h!]
\begin{center}
\begin{minipage}[h]{0.49\linewidth}
\includegraphics[width=6.6cm]{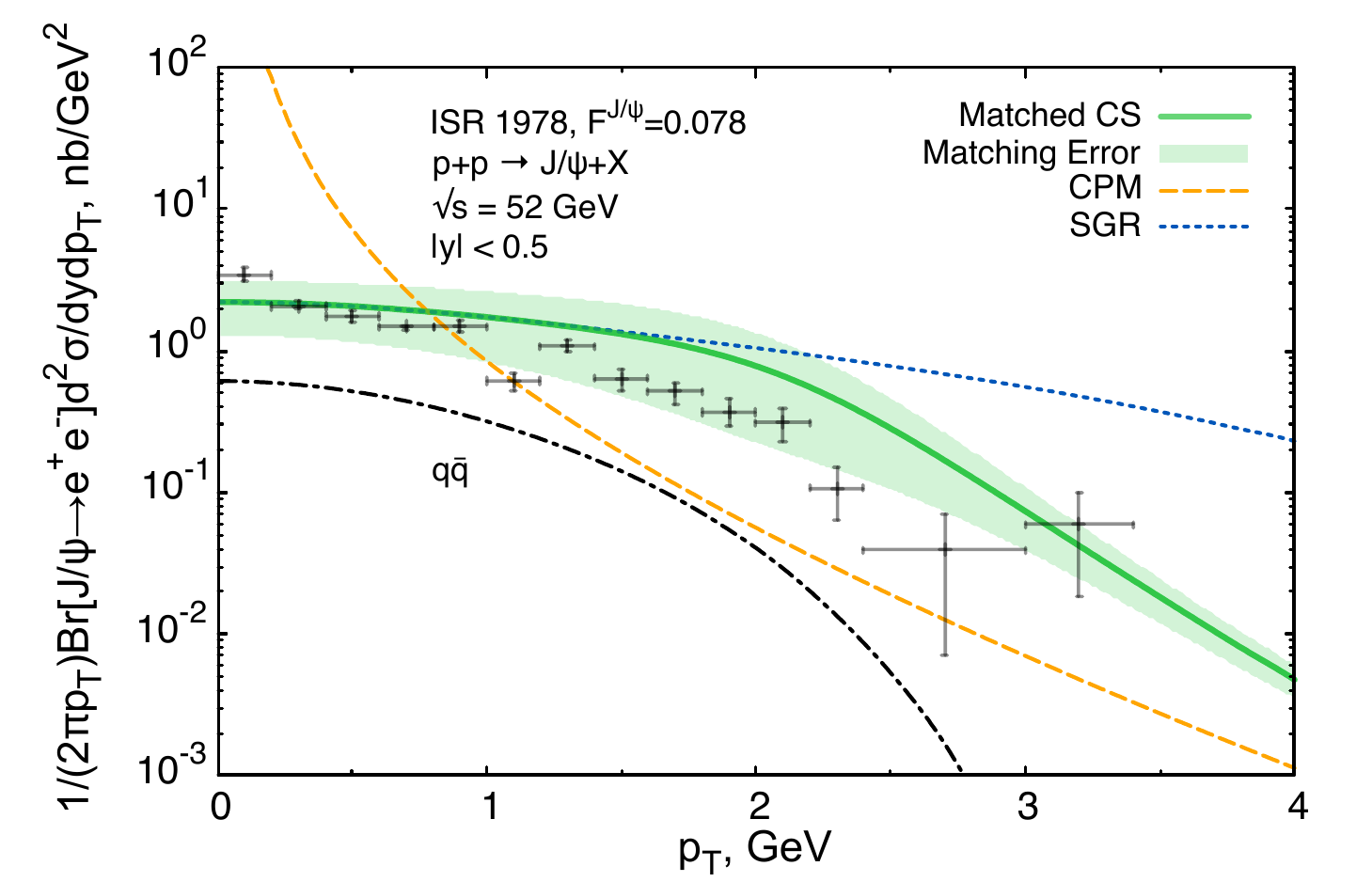}
\end{minipage}
\hfill
\begin{minipage}[h]{0.49\linewidth}
\includegraphics[width=6.6cm]{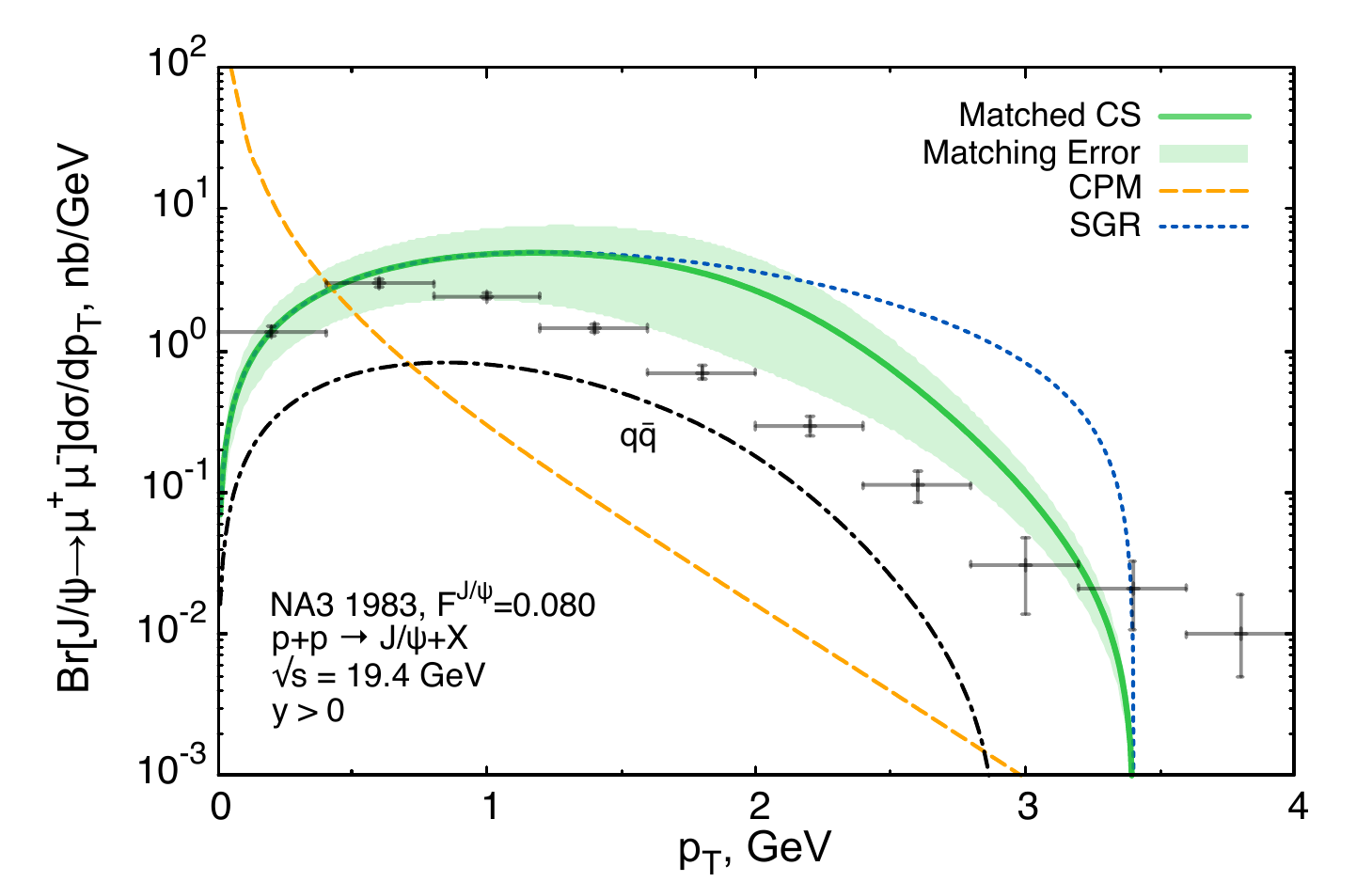}
\end{minipage}
\caption{The cross section of $J/\psi$ production versus transverse momentum at $\sqrt{s} = 52$ GeV (left panel) and $\sqrt{s} = 19.4$ GeV. The experimental data were taken from the ISR~\cite{Clark:1978mg} and NA3~\cite{NA3:1983ltt} Collaborations} 
\label{fig:6}
\end{center}
\end{figure}

\begin{figure}[b!]
\begin{center}
\begin{minipage}[h]{0.49\linewidth}
\includegraphics[width=6.6cm]{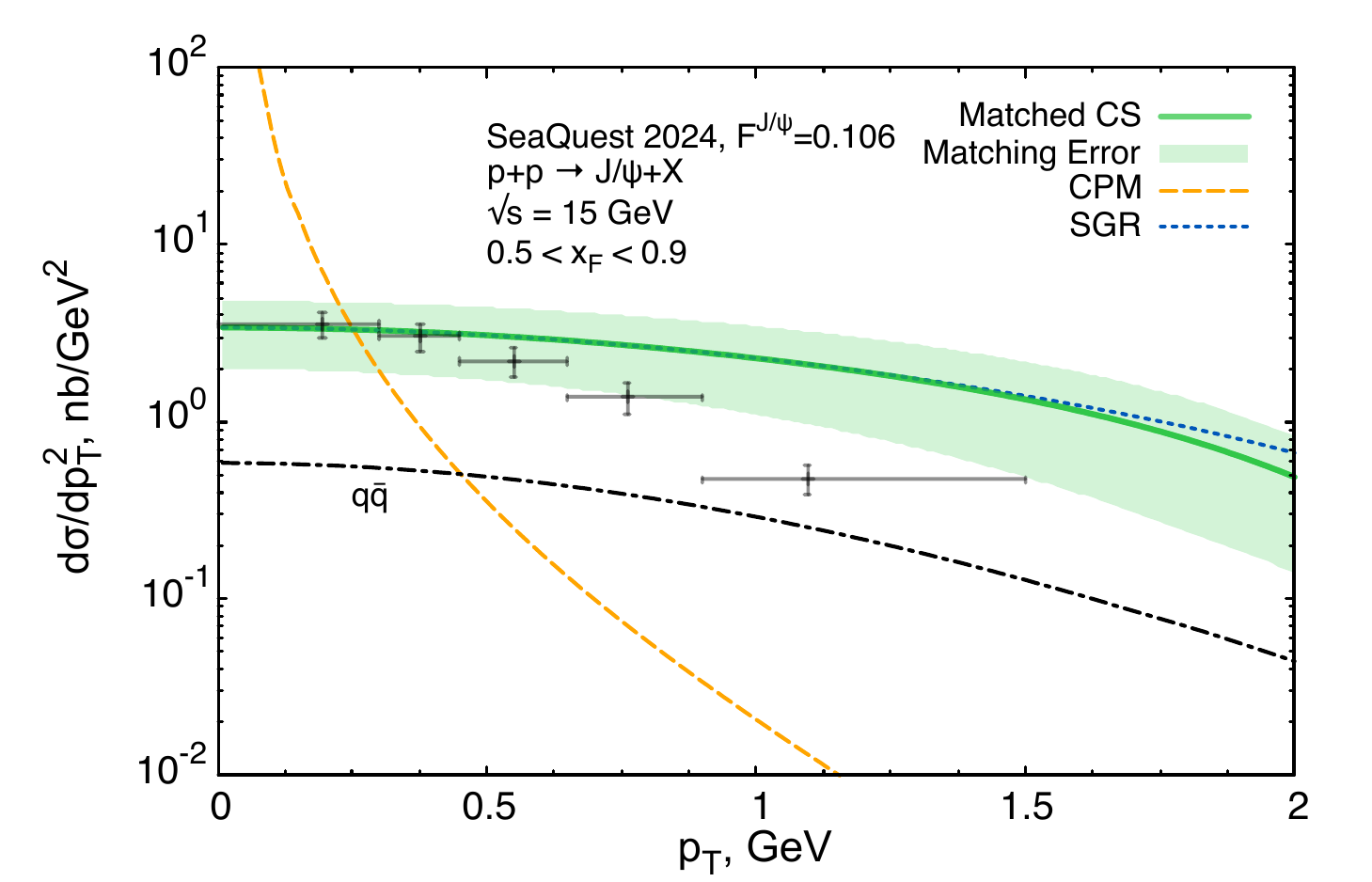}
\end{minipage}
\hfill
\begin{minipage}[h]{0.49\linewidth}
\includegraphics[width=6.6cm]{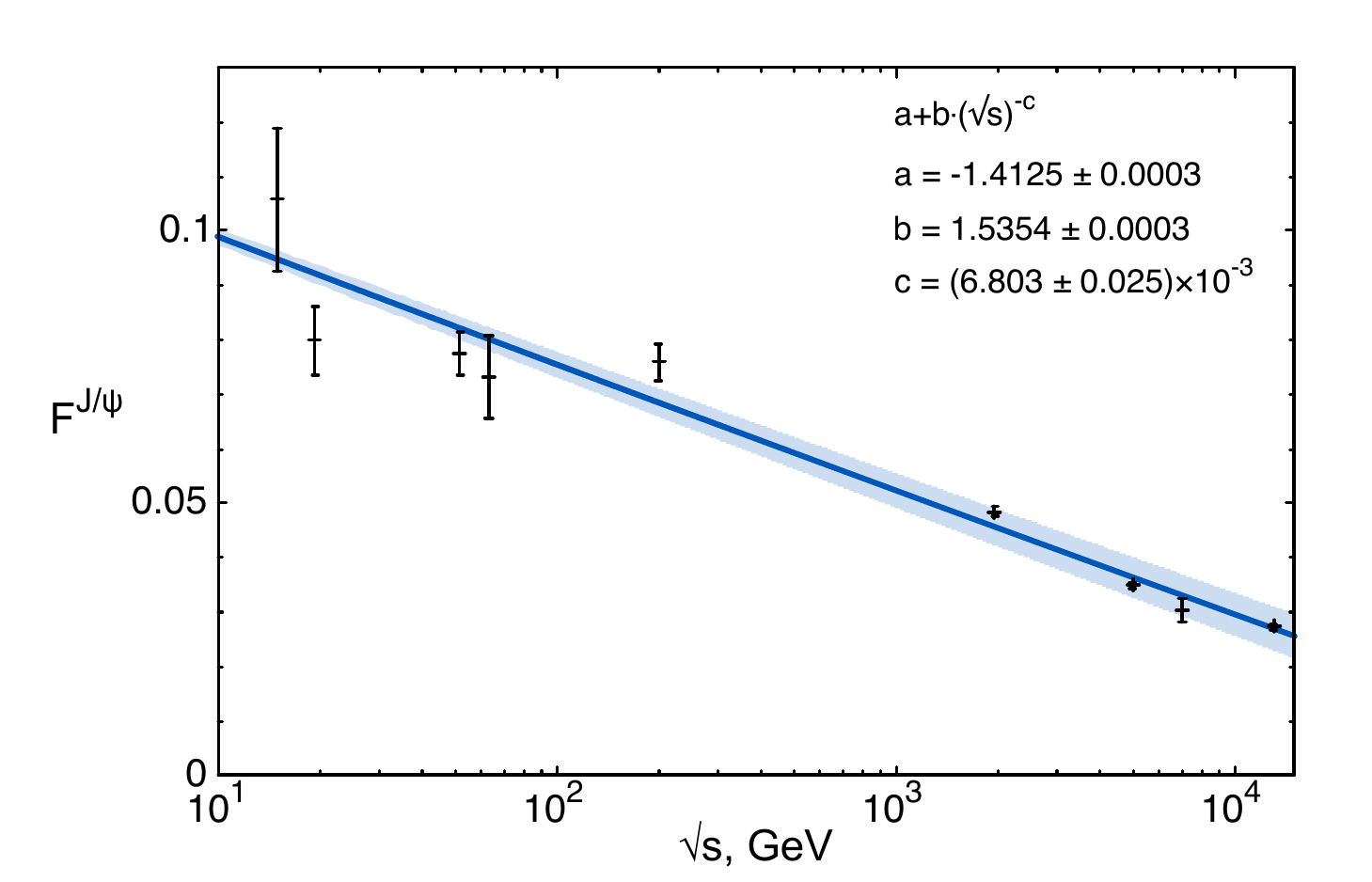}
\end{minipage}
\caption{The cross section of $J/\psi$ production versus transverse momentum at $\sqrt{s} = 15$ GeV (left panel), the experimental data were taken from the SeaQuest Collaboration.~\cite{SeaQuest:2024qdw} The result of $F^{J/\psi}$ fitting and approximation of obtained values (right panel)} 
\label{fig:7}
\end{center}
\end{figure}

Further, we provide our results on prompt $J/\psi$ production calculations within the approaches described in the previous sections. We made calculations for a wide range of center-of-mass energies $\sqrt{s} = 15$ GeV -- $13$ TeV,~\cite{LHCb:2015foc, CMS:2010nis, LHCb:2021pyk, CDF:2004jtw, PHENIX:2011gyb, Clark:1978mg, NA3:1983ltt, SeaQuest:2024qdw} our results are depicted on the Figs.~\ref{fig:1}--\ref{fig:7}. The SGR curves are blue and dotted, the fixed-order CPM curves are orange and dashed, green solid lines are for matched cross section, the uncertainties of matching are shown with the green bands around, the quark-antiquark annihilation subprocess contribution are also shown with black lines for comparison. We made a common fit of $F^{J/\psi}$ parameter within the TMD PM at $p_T^{} < 1$ GeV and the CPM at $p_T^{} > 5$ GeV for each $\sqrt{s}$ value and the summary of the series of fit procedures is shown on the right panel of the Fig.~\ref{fig:7}. As it is shown, the $F^{J/\psi}$ value increases when the center-of-mass energy $\sqrt{s}$ decreases. This dependency is fitted with the power-like function of the following kind \mbox{$F^{J/\psi} (\sqrt{s}) = a + b \cdot (\sqrt{s})^{-c}$} in order to use it for the SPD NICA experiment predictions. The parameters of the fitting function were found as \mbox{$a = -1.4125 \pm 0.0003$}, \mbox{$b = 1.5354 \pm 0.0003$}, \mbox{$c = (6.803 \pm 0.025) \cdot 10^{-3}$}, uncorrelated uncertainties correspond to one standard deviation, measurement units are omitted here.


\clearpage

\begin{figure}[b!]
\vspace{-0.4cm}
\begin{center}
\begin{minipage}[h]{0.49\linewidth}
\includegraphics[width=6.6cm]{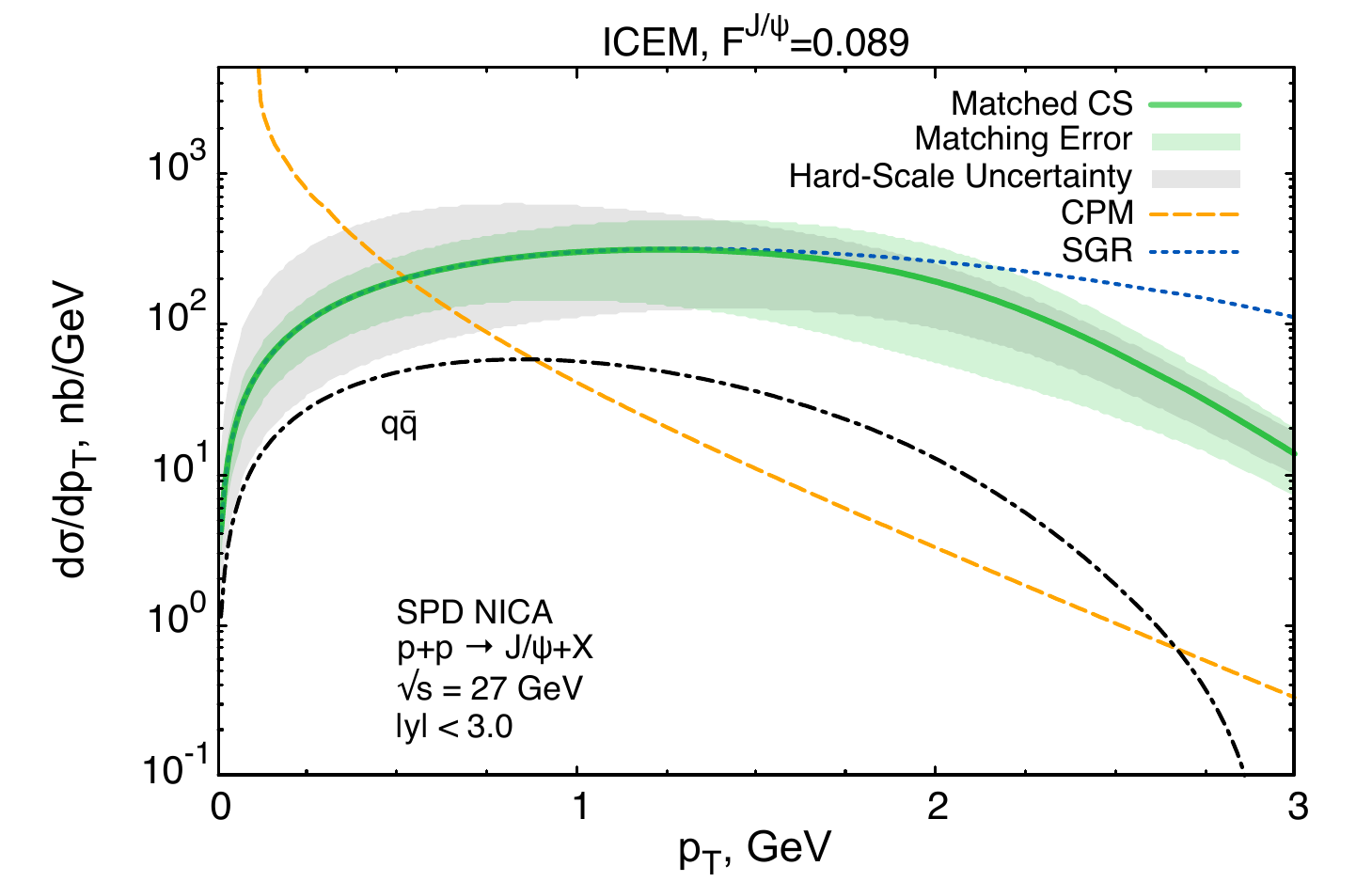}
\end{minipage}
\hfill
\begin{minipage}[h]{0.49\linewidth}
\includegraphics[width=6.6cm]{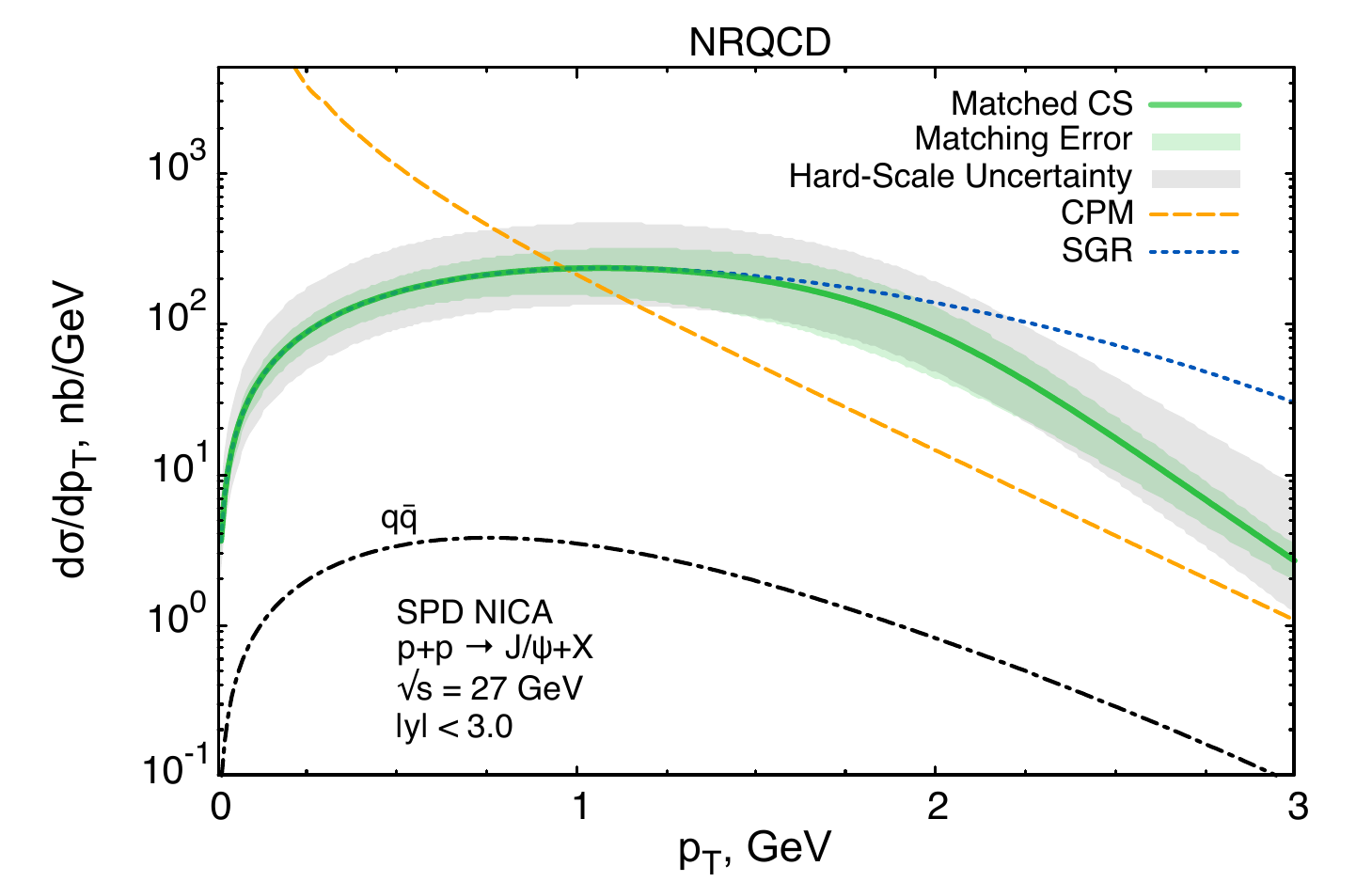}
\end{minipage}
\caption{The cross section of $J/\psi$ production versus charmonium transverse momentum at $\sqrt{s} = 27$~GeV for the SPD NICA experiment: within the ICEM (left panel) and the NRQCD (right panel) frameworks} 
\label{fig:8}
\end{center}
\end{figure}

As demonstrated on the plots, the curves of matched cross section agree satisfactorily with the experimental data, but at the intermediate-$p_T^{}$ domain one can observe some overestimation, though the predictions still agree with the data within the bands of matching uncertainty. The fixed-order CPM contribution behaves much worse at the lower center-of-mass energies of $15$ and $19.4$ GeV that is why, at these values of energy, only small-$p_T^{}$ region of calculations agrees with the data. 

The SPD NICA energy $\sqrt{s} = 27$ GeV lies between ranges of center-of-mass energy where the described combination of models works and does not, therefore we can provide predictions for SPD NICA with the value of $F^{J/\psi} = 0.089$ extracted from the fitting curve and at least compare it with our previous results~\cite{Saleev:2025ryh} obtained in the NRQCD framework. The prediction within two hadronization models and comparison of them are shown on the Fig.~\ref{fig:8}. Uncertainties from two sources are depicted on these plots: matching uncertainty in accordance with formula~(\ref{eq:uncertainty}) and uncertainty caused by variation of scales $\mu = \mu_R^{}$ with a factor of $2$. The agreement between the matched cross section curves within the uncertainties is quite good, but a discrepancy can be observed for the quark-antiquark subprocess contributions.

\section{Conclusion}

In the present paper, we study of prompt $J/\psi$ production in the Improved Color Evaporation model within the framework of SGR approach, CPM and InEW scheme. This combination of models allows to describe $J/\psi$ production spectrum at the arbitrary transverse momentum. 

Firstly, we made calculations in LL-LO approximation in the SGR approach and LO in the fixed-order CPM for available experimental data at center-of-mass energies $\sqrt{s} = 15$ GeV -- $13$ TeV. We obtained satisfying results for these data, except a crucial underestimation of the CPM contributions at the lowest energies. We fitted our calculations to the corresponding experimental data for each value of $\sqrt{s}$. The dependency of the ICEM phenomenological parameter $F^{J/\psi}$ on center-of-mass energy was shown. Secondly, we made predictions for kinematics of the SPD NICA forthcoming experiment at $\sqrt{s} = 27$ GeV. The predictions made in the present paper within the ICEM agree with those performed within the NRQCD of our previous study.~\cite{Saleev:2025ryh}

We do not expect significant differences between results of the current LO approximation in the strong coupling constant $\alpha_s^{}$ and more careful calculations in the NLO after redefining the phenomenological constant $F^{J/\psi}$ of the ICEM.

\section*{Acknowledgments}

The work is supported by the Foundation for the Advancement
of Theoretical Physics and Mathematics BASIS, grant No. 24-1-1-16-5,
and by the grant of the Ministry of Science and Higher Education of
Russian Federation, No. FSSS-2025-0003.

\section*{ORCID}

\noindent Vladimir Saleev - \url{https://orcid.org/0000-0003-0505-5564}

\noindent Kirill Shilyaev - \url{https://orcid.org/0009-0005-0531-883X}

\bibliographystyle{ws-mpla}
\bibliography{references}

\end{document}